\documentclass[conference]{IEEEtran}
\def\ARXIV{}\def\AUTHORS{}

\usepackage{tikz}
\usetikzlibrary{matrix}
\usetikzlibrary{calc}
\usepackage{amsmath}

\usepackage{mathptmx} %
\usepackage{fancyhdr}
\usepackage[normalem]{ulem}
\usepackage{flushend}
\usepackage{comment}
\usepackage{listings}
\usepackage{letltxmacro}
\usepackage{xparse}

\makeatletter
\@namedef{ver@lineno.sty}{9999/12/31}
\@namedef{opt@lineno.sty}{}
\makeatother

\makeatletter
\LetLtxMacro\ieeetran@appendix\appendix
\AtBeginDocument{%
\RenewDocumentCommand{\appendix}{o}{%
  \IfValueTF{#1}{%
    \ieeetran@appendix[#1]%
  }{%
    \ieeetran@appendix%
  }%
}
}
\makeatother

\ifdefined\PREPARXIV
\usepackage[finalizecache,cachedir=.,newfloat]{minted}
\else
\ifdefined\ARXIV
\usepackage[frozencache,cachedir=.,newfloat]{minted}
\else
\usepackage[newfloat]{minted}
\fi
\fi

\usepackage{hyperref}
\usepackage[nameinlink,capitalize]{cleveref}
\usepackage{color}
\usepackage{threeparttable}
\usepackage{booktabs}
\usepackage{tabularx}
\usepackage{algorithm}
\usepackage{algorithmicx}
\usepackage{algpseudocode}
\usepackage{mathtools}
\usepackage{pifont}
\usepackage{footnote}
\usepackage[compatibility=false]{caption}
\usepackage{subcaption}
\usepackage{paralist}
\usepackage{multirow}

\hypersetup{
  bookmarks=true,
  breaklinks=true,
  letterpaper=true,
  colorlinks,
  linkcolor=black,
  citecolor=blue,
  urlcolor=black
}

\newcommand\longvar[1]{\mathchardef\UrlBreakPenalty=100
\mathchardef\UrlBigBreakPenalty=100\url{#1}}

\def\pname{SPAM}
\def\pnameFULL{Stateless Permutation of Application Memory}
\def\pnameFULLEmph{\textit{Stateless} Permutation of Application Memory}
\def\pnameFULLSec{Stateless Permutation of Application\\ Memory (\pname)}

\def\b2p{Buf2Ptr}
\def\eg{e.g.,}
\def\ie{i.e.,}

\crefformat{figure}{#2Figure~#1#3}
\crefformat{table}{#2Table~#1#3}
\crefformat{listing}{#2Listing~#1#3}
\crefformat{algorithm}{#2Algorithm~#1#3}
\crefformat{footnote}{#2\footnotemark[#1]#3}
\AtBeginEnvironment{appendices}{\crefalias{section}{appendix}}

\newcommand{\cmark}{\ding{51}}
\newcommand{\xmark}{\ding{55}}
\newcommand{\cone}{{\large\ding{182}}}
\newcommand{\ctwo}{{\large\ding{183}}}
\newcommand{\cthree}{{\large\ding{184}}}

\newcommand\fakesec[1]{\vspace*{2pt} \noindent \hskip .01in \textbf{#1}}
\newcommand{\tabitem}{~~\llap{\textbullet}~~}
\newcommand{\pie}[1]{%
  \begin{tikzpicture}
    \draw (0,0) circle (1ex);\fill (1ex,0) arc (0:-#1:1ex) -- (0,0) -- cycle;
  \end{tikzpicture}%
}
\newcommand{\incode}[1]{{\mintinline{cpp}{#1}}}
\newcommand{\runtimefn}[1]{\textproc{#1}}

\newenvironment{algo}[2]{
  \begin{algorithm}[ht!]
    \small
    \caption{#1}
    \label{#2}
  }{
\end{algorithm}
  }

\ifdefined\AUTHORS
\IEEEoverridecommandlockouts
\fi

\begin{document}
\bstctlcite{IEEEexample:BSTcontrol}

\date{}

\title{\pname{}: \pnameFULL{}\vspace{-1em}}

\ifdefined\AUTHORS
\author{\IEEEauthorblockN{Mohamed Tarek Ibn Ziad\IEEEauthorrefmark{1}\thanks{\IEEEauthorrefmark{1}Both authors contributed
    equally to this work.},
Miguel A. Arroyo\IEEEauthorrefmark{1}, Simha Sethumadhavan}
\IEEEauthorblockA{Columbia University\\
New York, NY\\
Email: \{mtarek,miguel,simha\}@cs.columbia.edu}}
\fi

\maketitle

\begin{abstract}

In this paper, we propose the~\pnameFULLEmph{}~(\pname{}), a software defense that 
enables fine-grained data permutation for C programs. The key benefits include 
resilience against attacks that directly exploit software errors (\ie{} spatial and temporal memory 
safety violations) in addition to attacks that exploit hardware vulnerabilities such as ColdBoot, 
RowHammer or hardware side-channels to disclose or corrupt memory using a single
cohesive technique. Unlike prior work,~\pname{} is stateless by design 
making it automatically applicable to multi-threaded applications.

We implement~\pname{} as an LLVM compiler pass with an extension to the
\texttt{compiler-rt} runtime. We evaluate it on the C subset of the SPEC2017 benchmark 
suite and three real-world applications: the Nginx web server, the Duktape Javascript interpreter, 
and the WolfSSL cryptographic library. We further show~\pname{}'s scalability by 
running a multi-threaded benchmark suite. \pname{} has greater security coverage and
comparable performance overheads to state-of-the-art software techniques for
memory safety on contemporary \texttt{x86\_64} processors. Our security evaluation confirms~\pname{}'s
effectiveness in preventing intra/inter spatial/temporal memory violations by making 
the attacker success chances as low as~$\frac{1}{16!}$. 

\end{abstract}

\section{Introduction}\label{sec:introduction}

\begin{figure*}[!ht]
  \centering
  \includegraphics[width=0.9\textwidth]{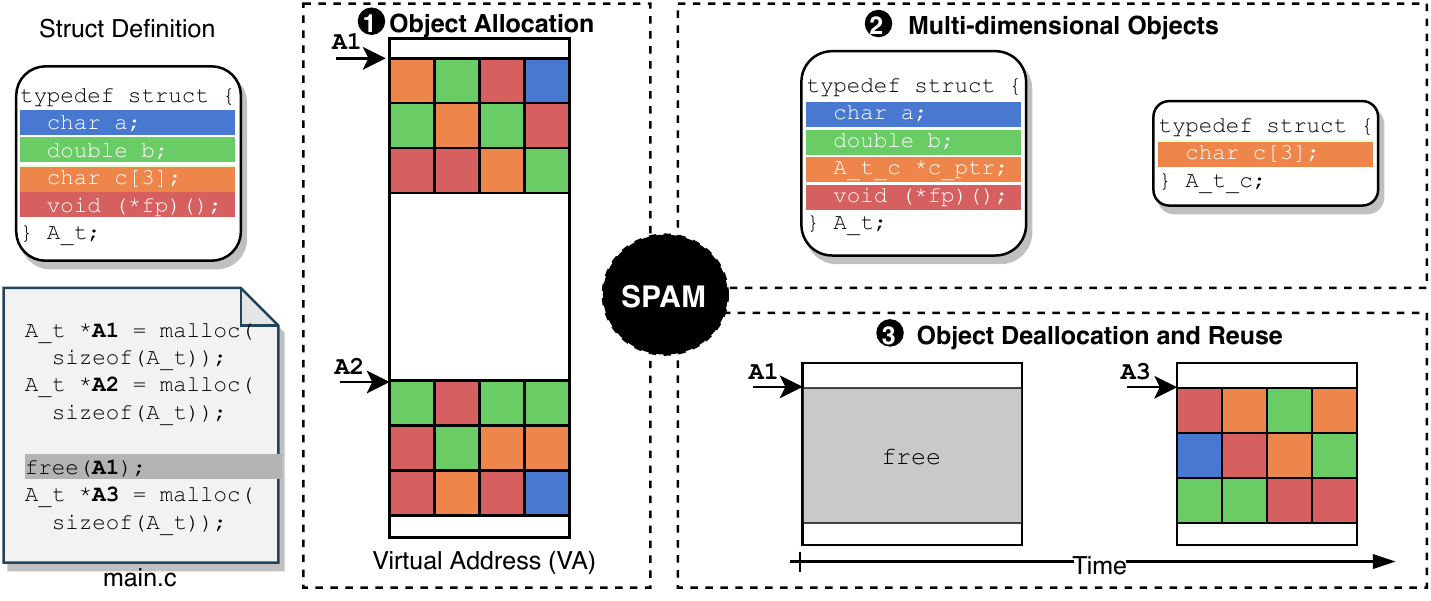}
  \caption{A sample program highlighting the lifecycle of an object in \pname{}:
    \cone{} every object has a unique permutation,~\ctwo{} buffer to
    pointer promotion (\b2p{}) for multi-dimensional objects and~\cthree{}
    deallocation of objects. 
  }\label{fig:permutations}
\end{figure*}

As reported by the Project Zero team at Google, memory corruption issues are the
root-cause of~$68\%$ of listed CVEs for zero-day vulnerabilities within the last
five years~\cite{projectzero2019:0day}. While current solutions can be
used to detect spatial and temporal software memory safety violations during
testing~\cite{Serebryany2012:ASAN, Duck2016:LowFatSW, Duck2018:effectiveSAN} 
and/or post-deployment~\cite{Nagarakatte2009:SoftBound, Nagarakatte2010:CETS, 
isomeron2015, Qualcomm2017, oleksenko2018intel, IntelCeT}, the recent development 
of hardware vulnerabilities
that can leak secrets (\eg{} hardware side-channels~\cite{Gruss2015:cache,
Marcus2017:side,Wenhao2017}, ColdBoot~\cite{Halderman2009:ColdBoot}) or corrupt
memory (\eg{} RowHammer~\cite{Kim2014:RowHammer}) need different specialized
approaches. The use of multiple security
countermeasures complicates the deployment of hardened software especially when
operational resources and budgets for security are limited. As pointed out by
Saltzer and Schroeder~\cite{Saltzer1975}, an economy of mechanism is valuable to
handle multiple software and hardware memory security issues.

In this paper, we present the \pnameFULL{} (\pname{}), a post-deployment 
defense that provides cohesive
protection against software and hardware memory corruptions with no explicit
metadata. To better understand how \pname{} works, let us consider the lifetime of an object from allocation to
deallocation as shown in~\cref{fig:permutations}.

\pname{} permutes the layout of data in program memory 
based on its location in the virtual address space. When a new region of memory is
allocated by the program,~\pname{} specifies a permutation for the data of this 
allocation. Thus, the mapping between memory instances (e.g., 
C structs) and their actual layout in memory is unknown to an attacker.
The allocation base address and size are used to derive one specific permutation 
out of~$\left(\frac{S}{G}\right)!$ different permutations, where~$S$ is the allocation size and~$G$ 
is the permutation granularity in number of
bytes.~\pname{} also uses a unique per-process key and a disclosure-resistant
pseudo-random number generator to mitigate record-and-reply attacks.
As illustrated in~\cref{fig:permutations}~(\cone), with \pname{} even if attackers 
have access to a memory safety vulnerability that
leads to an arbitrary read/write capability, they can no longer infer the order
or offsets of the victim data within the allocated region.

If the allocated object is multi-dimensional (\ie{} a compound data structure
with one or more buffer fields),
\pname{} converts them to one-dimension using a novel source-to-source transformation, \b2p{}, that promotes struct buffer 
fields into their own allocations. An example
transformation for a struct is shown in~\cref{fig:permutations}~(\ctwo). Here, the
buffer field \texttt{c} is promoted to a pointer \texttt{c\_ptr} that points to a
standalone allocation containing the original buffer field data. 
Both allocations,~\texttt{A1} and~\texttt{c\_ptr}, are permuted independently using 
their corresponding base addresses and sizes. In other words, \b2p{} reduces the
intra-object memory safety problem to
be equivalent to an inter-object one. 

When an object is freed the same memory region may be allocated for a new
object, posing a security concern. To mitigate this, \pname{} generates a
random value that is embedded within each pointer returned
by the memory allocator. Thus, a single memory region can have multiple different
permutations dependent on the random value. For example,
in~\cref{fig:permutations}~(\cthree{}) when~\texttt{A1} is freed, the memory allocator will use the
same memory again to satisfy the next allocation (i.e.,~\texttt{A3}). \pname{} embeds a different
random value in~\texttt{A3} resulting in a new permutation. As a result, even if the
same memory is allocated to a different object during the program
lifetime, the attacker has no guarantee that the data would be at the
same location as the freed object. Thus, \pname{}
provides complete coverage against software memory corruptions.

Additionally,
\pname{} provides resilience against hardware memory corruptions by
keeping the data permuted across the memory hierarchy (\ie{} caches and DRAM).
For example, sensitive data leaked by a ColdBoot attack is indistinguishable from
random as it would be permuted. Similarly, the randomness of physical memory
increases the complexity of a RowHammer attack as the attacker needs to know the
exact layout of adjacent data to decide where to trigger a bit flip. 
Moreover,~\pname{} provides a natural protection against speculative side-channels as
speculatively executed loads will always return permuted data making it harder
for an attacker to recover the original memory layout. More importantly, combinations of
techniques that aim to provide a similar level of protection to \pname{} may not
build on top of each other or may incur higher performance costs. As we argue in
this work, there is no single
cohesive solution that can address both software and hardware memory violations.

One important aspect of \pname{} is that it does not store
any metadata separately in protected memory
regions~\cite{Kim2019:Polar,Aga2019:Smokestack} or as part of the object
itself~\cite{ARM2019:MTE, Duck2018:effectiveSAN}. \pname{} dynamically
calculates permutations for every load and store. Thus, it neither introduces
additional storage overheads nor provides the opportunity to be manipulated by
an attacker. Especially in the context of multi-threaded programs, the lack of
metadata in \pname{} allows scalable performance of application code. In contrast, memory
safety techniques that rely on explicit metadata require proper synchronization to
maintain correct and secure behavior which negatively impacts scalability.

\pname{} is implemented within the LLVM compiler framework and currently targets
the \texttt{x86\_64} architecture. The experimental results show that~\pname{}
has comparable performance overheads to state-of-the-art software techniques for
memory safety on the C subset of SPEC2017 benchmark suite~\cite{Bucek2018:SPEC}, a web
server~\cite{nginx}, a Javascript interpreter~\cite{duktape}, and a cryptographic library~\cite{wolfssl}. 
We further show~\pname{}'s scalability by running a multi-threaded benchmark 
suite~\cite{Ahmad2015:CRONO}. Additionally, we
conduct a quantitative security analysis to demonstrate~\pname{}'s effectiveness
in preventing intra/inter spatial/temporal memory safety corruptions.

 The remainder of the paper is organized as follows. We specify the threat 
 model and assumptions in \cref{sec:threatmodel}. 
 We introduce~\pname{} in \cref{sec:systemoverview} and discuss our prototype
 implementation in \cref{sec:imp}. We then analyze the security of~\pname{} 
 in \cref{sec:securityanalysis}. \cref{sec:opt} highlights the main 
 performance optimizations, while \cref{sec:evaluation} extensively evaluates~\pname{} 
and compares it against state-of-the-art techniques.  
We summarize~\pname{} deployment considerations in \cref{sec:deploy} and 
discuss the current prototype limitations in \cref{sec:limitations}. 
\cref{sec:relatedwork} summarizes the related work. 
Finally, we conclude in \cref{sec:conclusion}.

\section{Threat Model \& Assumptions}\label{sec:threatmodel}

\fakesec{Adversarial Capabilities.}
We consider a powerful, yet realistic adversary model that is consistent with
previous work on software memory safety~\cite{Nagarakatte2009:SoftBound, Nagarakatte2010:CETS,
Duck2016:LowFatSW, Duck2018:effectiveSAN, Serebryany2012:ASAN, oleksenko2018intel}
with stronger assumptions against side-channel capable adversaries.
We assume that the adversary is aware of the applied defenses and has access to the source
code or binary image of the target program. Furthermore, the target program
suffers from a memory vulnerability that allows the adversary to
read from, and write to, arbitrary memory addresses.
We further assume that the attacker can
disclose information at run time~\cite{snow2013:JIT-CRA} and use
side-channels~\cite{Kocher2018:spectre} as part of the attack to read or manipulate memory contents.

\fakesec{Assumptions.}
~\pname{} requires the availability of source code for the target program.
This requirement is true for the majority of state-of-the-art techniques (see~\cref{tab:comparison}).
We also assume that the
underlying operating system (OS) enables W\textasciicircum{X}---i.e., no code
injection is allowed (non-executable data), and all code sections are
non-writable (immutable code).~\pname{} protection applies to all
instrumented code and libraries. Uninstrumented codes, such as third party
libraries and the operating system runtime, can be fully utilized without
limitations, but are not protected by~\pname{}. This model offers a path to incremental
adoption of \pname{}. Additionally, the \pname{} runtime is considered part of
the trusted-computing base (TCB). A number of low-overhead intra-process
isolation mechanisms can be utilized to harden the
runtime~\cite{ERIM:2019,Hodor:2019}. We leave the exploration
of runtime hardening for future work.
\section{\pnameFULLSec{}}\label{sec:systemoverview}

Our proposal to guarantee complete memory safety is to ensure that
the data layout is always permuted at a very
fine granularity. This permutation is meant to ensure that an attacker cannot
leak useful information or overwrite critical data within the program even in
the presence of memory safety vulnerabilities. To address our powerful threat
model, we require that information about permuted objects is neither
stored as part of the object itself, nor encoded in the binary version that can be accessed by an
attacker.

\subsection{Spatial Memory Safety} \label{subsec:spatial}

When a new memory region of size~$S$ is
allocated by the program, typically by calling \texttt{malloc},
\pname{} uses the base address ($BA$), size ($S$), and a $64$-bit per-process
key ($K$) to compute its permutation.\footnote{Without loss of generality, we
focus the discussion here about heap objects (C structs)
allocated by a memory allocator. In \cref{sec:imp}, we discuss how we
support stack \& globals.} As shown
in~\cref{fig:permutations}~(\cone), the struct \texttt{A\_t} is permuted
differently for each allocation instance (\ie{} \texttt{A1} and \texttt{A2}).
This new permutation defines the new offsets for the
individual data bytes within the struct. This runtime per-instance layout makes it
harder for attackers to construct a reliable exploit as they need to guess the
correct permutation.

\fakesec{Generating a permutation.} \pname{} is a format-preserving encryption (FPE)
scheme~\cite{DBLP:conf/fse/GranboulanP07,Rogaway10asynopsis} which uses the
\textit{Fisher-Yates shuffle}~\cite{FisherYates} algorithm (modernized
in~\cite{10.1145/364520.364540} and popularly known as the \textit{Knuth
shuffle}~\cite{10.5555/270146}) to produce an unbiased permutation (\ie{} every
permutation is equally likely)~\footnote{Prior work uses Fisher-Yates shuffle to
randomize allocations on the heap. Unlike \pname{}, which randomizes the layout
of each allocation independently, STABILIZER~\cite{Emery2013:STABILIZER} picks a
random base address for each new allocation from a pool of N addresses per each
size class, while MESH~\cite{Emery2019:MESH} uses shuffle vectors to perform
randomized allocation with low space overhead.}.To
generate \pname{} permutations we follow the approach described
in~\cref{alg:gen-perm}: (1) a PRNG is seeded with a per-process key, the
base address of an allocation, and the allocation size (2) the Fisher-Yates shuffle algorithm is then
executed to completion resulting in the set of permutations that are used to
access memory. \cref{alg:fisher} details the Fisher-Yates
shuffle algorithm which permutes an initialized array of elements in-place and returns the array.
The permutation generated by \pname{} can be viewed as a block
cipher, where cryptographically speaking, the strength of the scheme depends on
the PRNG. PRNGs only provide numbers in a fixed range. For Fisher-Yates, we
bound the PRNG's output using a modulo operation according to the number of
elements to be shuffled. While Fisher-Yates is unbiased, the modulo operation
may introduces bias affecting uniformity. We evaluate the uniformity of the
permutations generated by our implementation in \cref{subsec:rand-eval}.

\begin{algo}{Generating \pname{} Permutation}{alg:gen-perm}
  \footnotesize
  \textbf{Requires:} $G$---Permutation Granularity, $B$---Permutation Boundary\\
  \textbf{Inputs:} $K$---Per-Process Key, $BA$---Base Address, $S$---Allocation Size\\
  \textbf{Outputs:} $P$---Permutation

  \begin{algorithmic}[1]
  \Function{GenPerm}{$K$, $BA$, $S$}
  \State $R \gets$ \Call{PRNGSeed}{$K$, $BA$, $S$} \Comment{Initialize PRNG}
  \State $C \gets B/G$ \Comment{Define permutation chunks} 
  \State $P \gets$ \Call{Fisher-Yates}{$R$, $C$} \Comment{Get permutation}
  \State \Return $P$
  \EndFunction
  \end{algorithmic}
\end{algo}

\begin{algo}{Fisher-Yates Shuffle}{alg:fisher}
  \footnotesize
  \textbf{Inputs:} $R$---Random Number Generator, $C$---Permutation Chunks\\
  \textbf{Outputs:} $P$---Permutation

  \begin{algorithmic}[1]
    \Function{Fisher-Yates}{R,C}
    \State $P \gets$ \Call{Init}{C} \Comment{Initialize array of $C$ elements} 
    \For{$i$ from $C-1$ to $1$}
    \State $j \gets$ \Call{R}{ } $\% i$ \Comment{Generate number s.t. $0 \leq j
      \leq i$}
    \State \Call{Swap}{P[i],P[j]}
    \EndFor
    \State \Return $P$ \Comment{The permuted indexes.}
    \EndFunction
  \end{algorithmic}
\end{algo}

\begin{figure}
  \centering
  \includegraphics[width=0.45\textwidth]{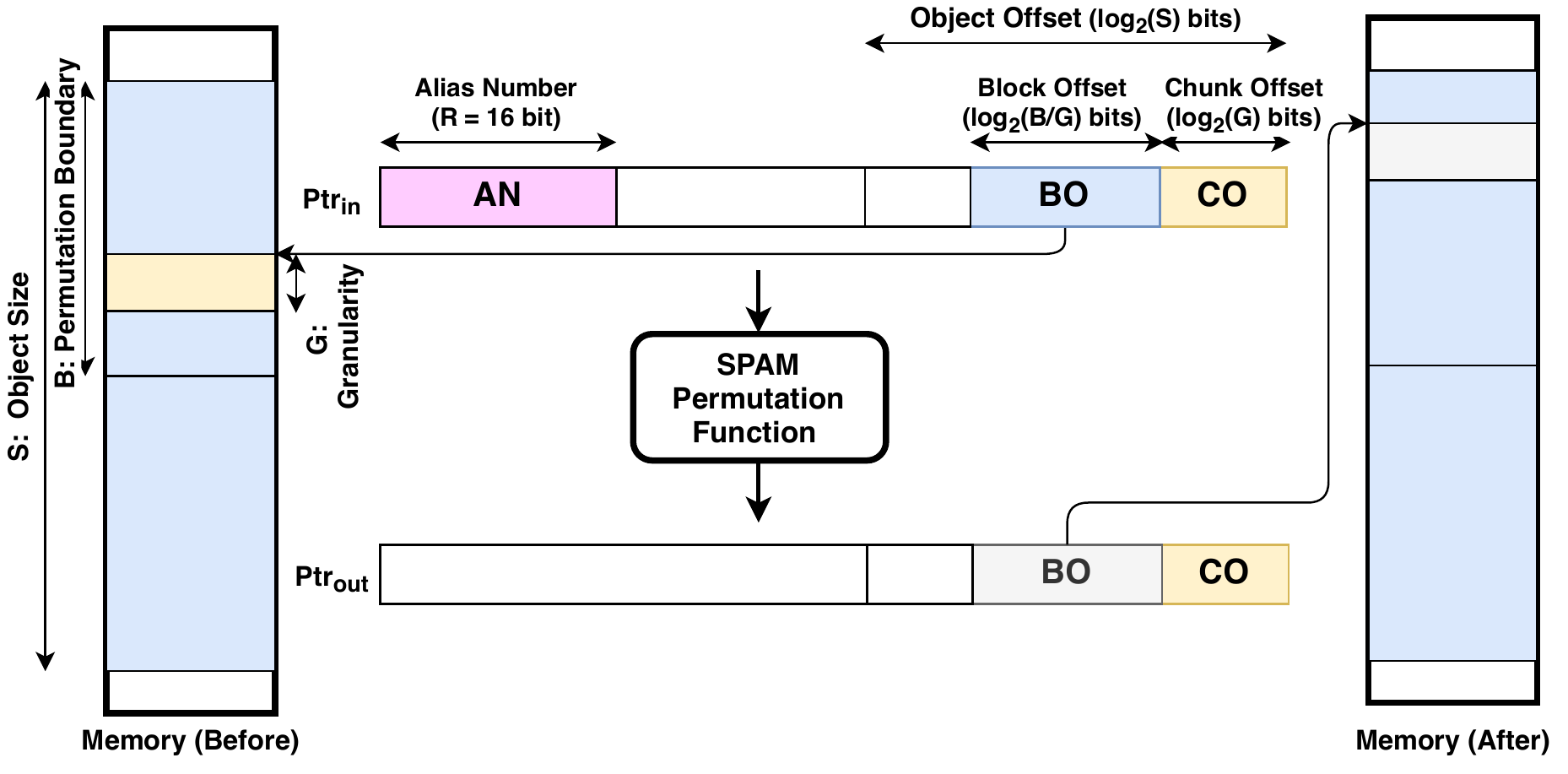}
  \caption{The process of updating pointers in~\pname{}.}\label{fig:perm-runtime}
\end{figure}

\begin{figure}
  \centering
  \includegraphics[width=0.45\textwidth]{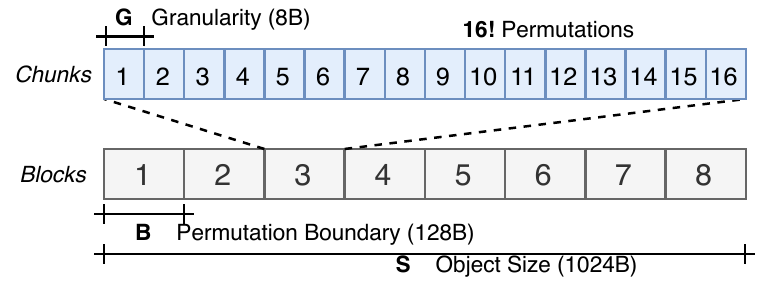}
  \caption{An example of how permutations are computed for an object of size,~$S = 1024$ bytes. We use a permutation boundry,~$B = 128$ bytes with a~$G = 8$ bytes permutation granularity, resulting in $\frac{B}{G} = 16!$ possible permutations.}\label{fig:perm-example}
\end{figure}

\fakesec{Updating pointers.} A new pointer to access the appropriate
memory location is computed given a pointer and
the provided permutation. This process is shown in
more detail in~\cref{fig:perm-runtime}.~\pname{} only changes 
the block offset bits that define the permuted location of a data chunk. 
The choice of permutation boundary, $B$, is
arbitrary, but defines the minimum allocation size.
For example as shown in~\cref{fig:perm-example}, an object of size~$S$ bytes will
consist of~$\frac{S}{B}$ permutation blocks, where each block consists of
$\frac{B}{G}$ chunks. Each block is then permuted given a per-process key, the object
base address and size. For objects greater in size than the permutation boundary,
we divide them into permutation blocks and permute them separately. 

\fakesec{(Un-)permuting memory.}
External library calls (\eg{} the ones made to
\texttt{libc}) are quite common. To maintain compatibility with uninstrumented external
code,~\pname{} provides~\runtimefn{unpermute} and~\runtimefn{permute}
primitives. The \runtimefn{unpermute} and \runtimefn{permute} primitives are
emitted for pointer arguments before and after external calls, respectively. 

\subsection{Sub-object Memory Safety} \label{subsec:sub}

To provide intra-object memory safety,~\pname{} proposes a novel application of an
idea called, \b2p{}, that has been previously used in the area of data layout optimizations 
for enhancing performance~\cite{Hundt2006:DLO,Roy2016:StructSlim,Yu2018:LWP}.
\b2p{} promotes array or
buffer fields defined in C/C++ structures to be independent, reducing the problem of
intra-object allocation to be equivalent to inter-object allocation. This allows
\pname{} to rely on the same security guarantees discussed in \cref{subsec:spatial}.
\b2p{} eliminates all security concerns about 
structs contiguously laid out in memory (\eg{}
\incode{malloc(10*sizeof(struct Foo))}) as they would no longer contain
arrays to overflow. To illustrate a \b2p{} transformation,
consider the example in~\cref{lst:b2p-example}. Array fields within a structure
are replaced with a promoted pointer (\eg{} \incode{p_buf}) and a new structure
containing the original array is defined (\eg{} \incode{Foo_buf}). As a result
of this transformation, allocations, deallocations, and usages of the original
field must also be properly promoted. For example, 
an allocation for a composite data type (\eg{} \incode{Foo}) becomes
separate allocations based on the number of fields promoted (\eg{}
\incode{Foo_buf}).

\begin{listing}[!h]\footnotesize
  \centering
  \begin{subfigure}[t]{0.22\textwidth}
    \centering
    \begin{minted}
      [
      linenos,
      fontsize=\footnotesize,
      mathescape,
      autogobble,
      stripnl=false,
      numbersep=1.5pt
      ]
      {cpp}
     
    struct Foo {
      char buf[10];
    };




    struct Foo *f = malloc(
      sizeof(struct Foo));



    f->buf[7] = 'A';

    free(f);
    
    \end{minted}
    \caption{Original}\label{lst:b2p-original}
  \end{subfigure}
  \begin{subfigure}[t]{0.22\textwidth}
    \centering
    \begin{minted}
      [
      fontsize=\footnotesize,
      mathescape,
      autogobble
      ]
      {cpp}
    // Promoted Type
    struct Foo_buf {
      char buf[10];
    };
    struct Foo {
      struct Foo_buf *p_buf;
    };
    // Promoted Allocations
    struct Foo *f = malloc(
      sizeof(struct Foo));
    f->p_buf = malloc(
      sizeof(struct Foo_buf));
    // Promoted Usages
    f->p_buf->buf[7] = 'A';
    // Promoted Deallocations
    free(f->p_buf);
    free(f);
    \end{minted}
    \caption{Transformed}\label{lst:b2p-transformed}
  \end{subfigure}
  \captionof{listing}{An example of~\b2p{} transformation.}\label{lst:b2p-example}
\end{listing}

\subsection{Temporal Memory Safety} \label{subsec:temp}

\begin{figure}[!t]
  \centering
  \includegraphics[width=0.45\textwidth]{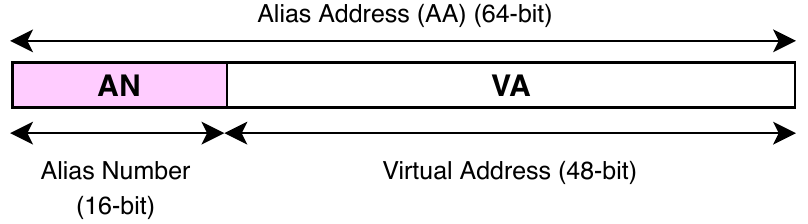}
  \caption{The construction of alias addresses created by~\pname{} to provide
probabilistic temporal memory safety.}\label{fig:AA}
\end{figure}

Permuting the object layout alone as described in \cref{subsec:spatial} is not
sufficient to prevent temporal attacks such as use-after-free.
This is primarily due to the fact that the attacker's object and the victim one
share the same base address (\ie{} due to the deterministic allocator behavior)
and key (\ie{} as both objects correspond to the same process). We observe that
the problem is due to address reuse among allocations. We avoid this problem by
re-purposing the currently unused high order bits of the virtual address (VA).
As shown in~\cref{fig:AA},~\pname{} randomly chooses a~$R$-bit alias number
(AN) and encodes it within the most significant~$R$-bits of the base address.
As~\pname{} generates permutations based on the entire address, named
alias address (AA), we get a new permutation for the same memory region (even if
the lower order bits of the VA remains the same).

\cref{fig:permutations}~(\cthree{}) shows how~\pname{} generates a new permutation for the same
object after freeing the old one. Each allocation can
have up to~$2^{R}$ different aliases (where $R$ depends on the hardware
architecture), with each alias having its own
permutation. \pname{} drops the AN bits whenever the alias address is passed to the
\incode{free} function so that the processor functionality
for address handling remains unaffected by our security modification.
\cref{alg:creation,alg:access,alg:delete} summarize
the steps needed for \pname{} to provide its spatial and temporal memory safety guarantees.

\begin{algo}{Allocation Creation}{alg:creation}
\footnotesize
\textbf{Requires:} $M$---Memory Allocator, $RNG$---Random Number Generator\\
\textbf{Inputs:} $S$---Allocation Size\\
\textbf{Outputs:} $AA$---Alias Address

\begin{algorithmic}[1]
  \Function{Alloc}{S}
  \State $VA \gets$ \Call{M}{S} \Comment{Returns Virtual Address}
  \State $AN \gets$ \Call{RNG}{ } \Comment{Generate 16-bit random number}
  \State $AA \gets \{AN, VA[47:0]\}$ \Comment{Assemble Alias Address}
  \State \Return $AA$
  \EndFunction
\end{algorithmic}
\end{algo}

\begin{algo}{Allocation Access}{alg:access}
\footnotesize
\textbf{Requires:} $K$---Per-Process Key, $G$---Permutation Granularity\\
\textbf{Inputs:} $AA$---Alias Address\\
\textbf{Outputs:} $VA_{new}$--Virtual Address
\begin{algorithmic}[1]
  \Function{Access}{$AA$}
  \State $BA \gets$ \Call{GetBasePtr}{$AA$} \Comment{Get allocation base address}
  \State $S \gets$ \Call{GetSize}{$BA$} \Comment{Get allocation size}
  \State $P \gets$ \Call{GenPerm}{$K$,$BA$,$S$} \Comment{Get permutation}
  \State $VA \gets$ \Call{Strip}{$AA$} \Comment{Strip Alias Number}
  \State $VA_{new} \gets$ \Call{GetPermPtr}{$VA$,$P$} \Comment{Get new address}
  \State \Return $VA_{new}$
  \EndFunction
\end{algorithmic}
\end{algo}

\begin{algo}{Allocation Delete}{alg:delete}
\footnotesize
\textbf{Requires:} $M$---Memory Allocator\\
\textbf{Inputs:} $AA$---Alias Address

\begin{algorithmic}[1]
  \Function{Delete}{$AA$}
  \State $BA \gets$ \Call{GetBasePtr}{$AA$} \Comment{Get allocation base address}
  \State $VA \gets$ \Call{Strip}{$BA$} \Comment{Strip Alias Number}
  \State \Call{Free}{$M$, $VA$} \Comment{$M$ frees memory normally}
  \EndFunction
\end{algorithmic}
\end{algo}

\section{Implementation}\label{sec:imp}

\begin{figure}
  \centering
  \includegraphics{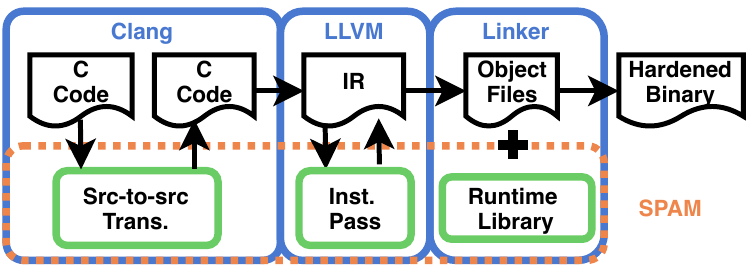}
  \caption{The \pname{} compiler infrastructure framework.}\label{fig:frame}
\end{figure}

\cref{fig:frame} provides an overview of our~\pname{} framework.
In this section, we discuss the different components.

\fakesec{Source-to-Source Transformation.}
We implement \b2p{} as a source-to-source transformation pass using Clang's
rewriter interface. The pass performs two main traversals over the AST. The
first traversal analyzes each translation unit to collect a whole program view
of composite data types (\eg{} structs) and their usages. This information is used
to determine what can be legally promoted, as explained in \cref{subsec:sub}. The
second traversal performs the actual rewriting.

\fakesec{Instrumentation Pass.} To handle heap memory, we implement an instrumentation 
pass at the LLVM IR level augmenting all the necessary loads and stores to invoke the
runtime, which computes accesses to the appropriate permuted memory locations.
We iterate over all loads and stores and emit a call
to the runtime to resolve the permuted memory location to be accessed.
Then, we determine whether calls refer to externally linked functions or
functions defined in other translation units. Finally, external
calls are broken down into two categories \emph{wrapped} and \emph{unwrapped}.
Wrapped calls are simply replaced with \pname{} specific variants that operate
on permuted memory to improve performance. Unwrapped calls are guarded by a pair of
\textproc{Unpermute} and \textproc{Permute} runtime functions if pointer
arguments are used in the relevant call. These operations ensure compatibility
with uninstrumented code by unpermuting/permuting data before/after executing 
uninstrumented code.~\cref{tab:wrappers} (in \cref{sec:opt-appendix}) 
includes a list of~\pname{} wrappers.

Our current prototype support for global and stack memory builds on top of
LowFat~\cite{duck2018extended,duck2018stack}. In addition to the
features provided by LowFat, for constant globals (\ie{} those typically in the
\incode{.data} section) we emit a call to a
\runtimefn{RegisterGlobal} runtime function. The \runtimefn{RegisterGlobal}
function is appended to the global constructors (\ie{} \incode{.ctor}) to
permute each global variable on
program load. Similarly, for stack variables usually passed by the OS (\eg{}
\incode{argv}), we emit a call to the \runtimefn{RegisterStack} runtime function
in order to permute memory on program start (\ie{} \incode{main}).

\fakesec{Runtime.}
\pname{}'s runtime,
implemented as an extension to LLVM's \texttt{compiler-rt}, encompasses the
following functionality: (1) getting an allocation's base address (2) looking up
the allocation's size (3) generating a permutation (4) calculating the memory
location to be accessed and (5) unpermuting and permuting memory for
compatibility with uninstrumented code. Our current implementation
builds on top of the LowFat~\cite{Duck2016:LowFatSW} allocator. Support for
other allocators is possible as long as they provide (1) and (2) making our
approach allocator agnostic. \cref{sec:add-runtime-api} provides an overview
of the runtime API.

\fakesec{Permutation Parameters.} In our current prototype, we use a permutation
granularity of~$8$B (i.e., every successive~$8$B in memory will remain
unpermuted while an entire~$8$B chunk can be stored in any place within a
permutation boundary). The main reason for this choice is that~$8$B accesses are
common on~$64$-bit systems. Using a smaller chunk size, while supported by our
implementation, will add additional overheads. We use a permutation boundary
of~$128$B to match the cache line size of the Last Level Cache~\footnote{Permutation
boundary can be easily tuned to match the requirements of different processors
(e.g.,~$64$B LLC cachelines on recent AMD processors).}. Our current prototype
targets the \texttt{x86\_64} architecture, which can have a maximum alias number
of $16$-bits as the virtual address consumes $48$-bits of the total $64$-bit
address space.

\section{Security Analysis}\label{sec:securityanalysis}

In this section, we first discuss the security guarantees 
provided by~\pname{} then we evaluate them quantitatively.

\begin{table}[!h]
\caption{The basic primitives that define a program and how~\pname{} affects them.}\label{tbl:primitives}
\resizebox{0.47\textwidth}{!}{%
\begin{tabular}{@{}lrl|rl@{}}
\toprule
                                 & \multicolumn{2}{c}{\textbf{Uninstrumented}} & \multicolumn{2}{|c}{\textbf{\pname{}}} \\ \midrule
\multirow{6}{*}{\textbf{Memory}} & \texttt{addr} $\leftarrow$  & \texttt{malloc(S)} & \texttt{baddr} $\leftarrow$ & \texttt{\pname{}\_malloc(S)}        \\
                                 &                &                            & \texttt{perm} $\leftarrow$ & \texttt{genperm(K, baddr, S)}   \\
                                 &                &                            & \texttt{paddr} $\leftarrow$ & \texttt{getaddr(addr, perm)}    \\
                                 &                & \texttt{free(addr)} &          & \texttt{\pname{}\_free(paddr)}      \\
                                 & \texttt{val} $\leftarrow$ & \texttt{load(addr)} & \texttt{val} $\leftarrow$ & \texttt{load(paddr)}            \\
                                 &                & \texttt{store(addr, val)} &          & \texttt{store(paddr, val)}      \\ \midrule
\textbf{Compute}                 & \texttt{val} $\leftarrow$ & \texttt{arith(val)}       & \texttt{val} $\leftarrow$      & \texttt{arith(val)}             \\ \midrule
\textbf{Control-Flow}            &                & \texttt{branch(addr)}               &          & \texttt{branch(addr)}           \\ \bottomrule
\end{tabular}%
}
\end{table}

\cref{tbl:primitives} shows the primitives that define a program. \pname{}'s mechanism for protecting memory involves
introducing new secure primitives that act as a shell to guard memory operations
as introduced in \cref{sec:systemoverview}. The permutation shell around the
memory operations guarantees that input and output operations that attackers can
use to subvert/control programs have uncontrollable behavior. Memory operations
can be used in two modes: relative or absolute. With absolute
read/write capability, an attacker can control the \texttt{addr} supplied to a
\texttt{load} or \texttt{store}. A relative read/write capability returns/updates a
value at an arbitrary offset from a known address (\eg{}
\incode{f->buf[7]} where $7$ is an example of an attacker controlled offset).
An overflow in 
the buffer field,~\incode{buf}, can corrupt other fields in 
object,~\incode{f}.

\pname{} revokes absolute capabilities as now every load/store instruction depends on a secret permutation
(\texttt{perm}) that is
not available to the attacker. This permutation is derived using a secure key (\texttt{K})
and is computed at runtime (using \texttt{genperm} and \texttt{getaddr}) for
every access. Alternatively, an attacker can read/write memory via an
external mechanism (\eg{} side-channels) to bypass the secure primitives.~\pname{} 
nullifies this external capability by keeping data permuted across the
entire memory hierarchy. This makes it impractical for an 
attacker to recover the memory contents. In addition to that,~\pname{} revokes relative 
capabilities using \b2p{} which reduces them into
absolute read/write capabilities. 

Common exploits build upon the absolute/relative read/write capabilities as
discussed above. We briefly describe how these exploits are each
specifically handled by \pname{}.

\subsection{Resilience to Common Exploits} \label{subsec:Qual}
\fakesec{Buffer under-/over-flows.}
\pname{} can defend against the exploitation of buffer overflows (and
underflows) by hiding the mapping between program objects and their actual layout
in memory. Even if the attacker has access to the source code and/or binary
image of the victim program, they cannot infer the layout of the victim object.
The same object can have multiple layouts based on its location in memory. With
high probability, with \pname{} in place, the attacker cannot corrupt or leak
information that can be used to mount many exploit variants. 
\pname{}'s protection applies to both inter- and intra-object safety (as \b2p{}
reduces the intra-object problem to inter-object).

\fakesec{Use-after-frees.}
As described in \cref{subsec:temp}, \pname{} provides temporal memory
safety via the alias address space. The same allocated virtual/physical memory
region can have up to~$2^{16}$ different aliases (each alias having its own
permutation). This alias address space is sufficient to
thwart use-after-free attacks, in which the type of the freed object 
aligns with the confused type of the new object, having a big impact on the
reliability of these exploits~\cite{projectzero2015:confusion}.
\textit{Heap Feng Shui} attacks exploit a memory allocator's determinism to arrange memory
so that it is favorable for an attacker to manipulate a victim
allocation~\cite{Sotirov2007:heapfengshui}. Similarly,
\pname{} relies on the alias address space to generate multiple permutations for the
same memory region, making it impractical to infer any information about an object 
layout based on another object, even if both have the same type. 

\fakesec{Uninitialized Reads.}
While \pname{} does not explicitly zero out memory that may have held security
sensitive data, the fact that the memory is left permuted after a free is
sufficient in many cases. The next program to use the same memory region 
will be assigned a different permutation key by default, making it impractical to recover 
the data by mistake. Additionally, an attacker trying to access this sensitive data would
need the appropriate permutation in order to unscramble the memory. The
same applies when peeking at memory with a variadic function misuse attack~\cite{CWE:FormatString}.

\fakesec{Control-Flow Hijacking and Data-Oriented Attacks.} Given a memory 
error, attackers can gain arbitrary memory read/write primitives. Attackers can 
then leverage such primitives to launch different attacks, such as control-flow 
hijacking~\cite{Bletsch2011:JOP, snow2013:JIT-CRA, Schuster2015:COOP, 
vanderVeen2017:10Years}, information leakage~\cite{Raoul2009:Leak}, or 
data-only attacks~\cite{Chen2005:Non-control, Hu2016:DOP, Pewny2019:STEROIDS, 
Cheng2019:SoKDOP}.~\pname{} effectively mitigates all of those attacks as it 
makes it harder for the attacker to utilize the memory read/write primitives. 
For instance, it is impractical to hijack the control-flow of the program 
(e.g., by overwriting a function pointer in a C struct) if the whole struct is 
permuted with~$\left(\frac{S}{G}\right)!$ different permutations. 
The same argument holds even for the more 
critical data-only attacks that corrupt the program without changing its 
control-flow.~\pname{} provides sufficient probabilistic guarantees for 
protecting the security-critical data structures of a program.    

\fakesec{Memory errors in uninstrumented code.} \pname{} unpermutes the data
that is passed to uninstrumented code (\eg{} library functions) while the rest
of the program data remains permuted. So, if the uninstrumented code has a
memory vulnerability it may only reliably corrupt the portion of data that is
passed to it. Unlike other techniques that provide no security guarantees for
uninstrumented code, \pname{} reduces the attack surface by keeping the rest
of the program data permuted.

\begin{listing}[!h]
  \centering
  \begin{minipage}{0.44\textwidth}
    \begin{minted}
      [
      linenos,
      fontsize=\footnotesize,
      mathescape,
      autogobble,
      stripnl=false,
      numbersep=3pt
      ]
      {cpp}
      if (i < sizeof(a)) { // mispredicted branch
        secret = a[i];
        val = b[64 * secret]; // secret is leaked 
      }
    \end{minted}
  \end{minipage}
  \caption{Example speculative execution attack.}\label{lst:spec-exec}
\end{listing}

\fakesec{Speculative Execution Attacks.}
With~\pname{}, utilizing
speculative exploits is more challenging for an attacker. Not only is the data
permuted, but the speculative (instrumented) load additionally uses a different permutation to
access the permuted data. Consider an attacker that tries to speculatively load
the secret value \incode{a[i]} using an out-of-bound index,~\incode{i}, as
shown in~\cref{lst:spec-exec}. In this case, they will end up with an
unpredictable value in \incode{secret} due to \pname{}'s security primitives
which permute the address of \incode{a[i]}.

\fakesec{Hardware Memory Violations.}
Let us consider an attacker wants to leak a
function pointer from a struct that has ten other fields. Using a side-channel,
the attacker leaks the entire struct. However, due to~\pname{}'s permutation
they would not be able to recognize the needed function pointer (or even reconstruct the
struct layout)\footnote{One may argue that using permutation granularity, $G = 8$ bytes, 
might help the attacker distinguish between pointer and non-pointer data items in a 
leaked struct. This can be thwarted by configuring our framework to use smaller 
values for $G$ in such cases.}. By the same principal, \pname{} provides indirect
protection against other types of
attacks such as RowHammer and ColdBoot attacks.

\fakesec{Chosen Data Attacks.}
One natural question to ask is whether an attacker can use data from a
computation to recover the data structure layout and therefore gain an insight into the
permutation. Let us assume that an attacker can inject $N$ unique values into $N$
fields of a struct. Let us further assume that the attacker can read/leak the
permutation. This read will give the attacker a permutation that is valid only
for the address allocated to the instance of the struct that has been primed
with unique known values. A different allocation (address) will have a different
permutation. To be able to predict the permutation for any address, the attacker
would have to do \texttt{AES-inverse(key, alias|address)} based on several reads
of permuted locations. The complexity of this attack is the complexity of
reversing AES with chosen plaintexts. Alternatively, an attacker may perform
\textit{Heap Feng Shui}~\cite{Sotirov2007:heapfengshui} style exploit and place
a similar vulnerable data structure at a location for which they know the
permutation. To carry out this attack, the attacker has to make at least
$2^{16}$ attempts because the alias number is chosen randomly for each
allocation at the same address.

\subsection{Security Litmus Tests} \label{subsec:Quan}

Here, we quantitatively evaluate the security guarantees provided by~\pname{} 
and compare it against two state-of-the-art techniques: AddressSanitizer (ASAN) 
and Intel MPX, as representatives of pre- and post-deployment memory safety 
solutions, respectively.

\begin{savenotes}
  \begin{table}
    \centering
    \caption{RIPE Results}\label{tab:ripe}
    \footnotesize

      \begin{tabular}{@{}llll@{}}
        \toprule
                         & \textbf{ASAN} & \textbf{MPX} & \textbf{SPAM} \\ \midrule
        \textit{Working Attacks} & 2/20          & 0/20         & 0/20          \\
        \bottomrule
      \end{tabular}%
  \end{table}
\end{savenotes}

\fakesec{RIPE.} We quantitatively evaluate security by using RIPE~\cite{Wilander2011:RIPE}, an open
source intrusion prevention benchmark suite. We first ported RIPE to 64-bit systems and compiled it
with~\pname{}. While
RIPE can support attacks on stack, globals and the heap, we focus solely on
heap-related attacks due to its additional temporal nature. For our baseline, the 
total number of attacks that survive with a native (non-protected) GCC and Clang 
is~$20$ attacks. The results for all tools are
summarized in~\cref{tab:ripe}. As expected, MPX\footnote{With \texttt{BNDPRESERVE=1} \& \texttt{-fchkp-first-field-has-own-bounds}\label{foot:mpx}} and \pname{} are able to prevent
all attacks due to providing intra-object protection. ASAN overlooks two
intra-object attacks in which a function pointer is corrupted. 

\fakesec{Microbenchmarks.} In addition to RIPE, we implement a small set of
security microbenchmarks. While RIPE tests for control-flow hijacking attacks,
our set of microbenchmarks aims to provide wider coverage. Thus, our set tests
against equally important categories of attacks, such as type-confusion and
information leakage. The complete set of results denoting the ability for a
given tool to detect specific vulnerabilities is summarized
in~\cref{tab:micro-security}. The results highlight the scope of each tool.
\emph{\pname{} is the only solution of those evaluated that is comprehensive
enough to provide coverage over the spectrum of these vulnerabilities.}

\begin{savenotes}
\begin{table}
  \centering
  \caption{Security Microbenchmarks}\label{tab:micro-security}
  \footnotesize
    \begin{tabular}{@{}llll@{}}
      \toprule
                                                          & \textbf{ASAN} & \textbf{MPX}$^{\ref{foot:mpx}}$ & \textbf{\pname} \\ \midrule
      \begin{tabular}[c]{@{}l@{}}Intra-Overflow\end{tabular} &   \xmark      & \cmark    &  \cmark    \\
      \begin{tabular}[c]{@{}l@{}}Inter-Overflow\end{tabular} &   \cmark      & \cmark    &  \cmark    \\
      \begin{tabular}[c]{@{}l@{}}Use-after-free\end{tabular} &   \cmark      & \xmark    &  \cmark    \\
      \begin{tabular}[c]{@{}l@{}}Type Confusion\end{tabular} &   \cmark      & \cmark      &  \cmark    \\
      \begin{tabular}[c]{@{}l@{}}Buffer Over-read\end{tabular}     &   \cmark      & \cmark    &  \cmark \\
      \begin{tabular}[c]{@{}l@{}}Uninitialized Read\end{tabular}  &   \xmark      &  \xmark    &  \cmark \\ \bottomrule
    \end{tabular}

    \vspace{1em}
\end{table}
\end{savenotes}

\begin{figure}[!ht]
  \centering
  \includegraphics[width=0.35\textwidth]{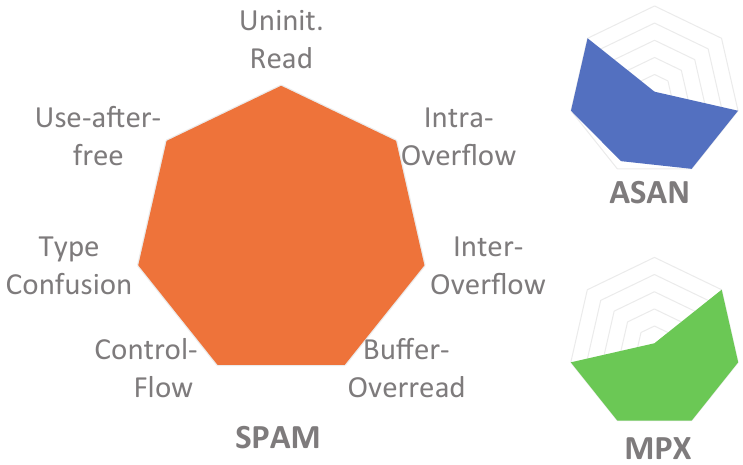}
  \caption{Quantitative security trade-offs for different tools.}\label{fig:cohesive}
\end{figure}

\fakesec{Cohesiveness.}\label{subsec:cohesive} We evaluate the security trade-offs between the
different tools. We use the security microbenchmarks and RIPE (shown as
Control-Flow) as indications of a tool's security coverage. 
The results are summarized in~\cref{fig:cohesive} with the overall area
of the polygon indicating the number of categories covered (\ie{} the larger the
area the greater the coverage). As we argue in this work, no
single tool can provide a cohesive solution for memory violations as \pname{} highlighting
the notability of its approach. In \cref{sec:relatedwork}, we discuss additional
hardware security related aspects of \pname{} in comparison to other
state-of-the-art techniques further highlighting its cohesiveness.

\section{\pname{} Performance Optimizations}\label{sec:opt}

In this section, we explain a set of optimizations that we enable in our current prototype 
for enhancing its performance.\footnote{The main focus of 
the current prototype is security coverage. We leave investigating other performance 
optimizations to future work.} 

\fakesec{Permutation Deduplication.} To avoid repeating the work of generating
permutations multiple times, at the beginning of a function we
generate the permutations for any pointer that is passed as a function input.
We propagate a permutation across multiple loads and stores (if they share the same base address)
instead of recomputing it. We pack this permutation
into an integer of an appropriate size depending on the permutation granularity
to efficiently pass it around. This integer is taken as an argument to our 
runtime function (\runtimefn{GetPermPtr}) to avoid re-generating a permutation for each
load/store. Similarly, for pointers defined later in the function's scope, 
a call to our generate permutation function is emitted and the packed integer
holding the permutation is propagated.

\fakesec{Software Caching.} Since all accesses to the same object 
have the same base address, we
exploit this locality by using a small lookup cache before accessing the
permutation function. We implement a direct-mapped software cache in 
our runtime to store the corresponding permutation for a given base 
address.\footnote{Our current prototype uses a
default cache size of~$2^{17}$ entries. Each entry stores~$64$-bit address
and~$64$-bit packed permutation, resulting in a total size of~$2$MB.}

If this optimization is enabled, an attacker may manipulate the 
cache contents to undermine~\pname{} security. To handle this issue, we propose 
two protection mechanisms. The first solution is to store the cache itself in a permuted 
fashion much like the rest of the application memory. In this case, the load time
address of the cache will act as the base address. Alongside with the
per-program key, a unique permutation is generated for the cache contents.
Then, all cache accesses from the~\pname{} runtime are updated to use it
directly with no additional latency. Alternatively, if available,
functionality like Intel Memory Protection Keys (MPK)~\cite{Park2019:libmpk,ERIM:2019} can be used to
protect the cache. MPK provides a single, unprivileged instruction,
\texttt{wrpkru}, that can change page access permissions by re-purposing four
unused bits in the page table. First, the cache is placed in pages that have a
particular protection key, forming the sensitive domain. Then, all cache
accesses from the \pname{} runtime are guarded with the \texttt{wrpkru}
instruction so that no other loads/stores have access to the cache contents.
For completeness, we evaluate MPK cache protection overheads on an Amazon Web Services (AWS)
\texttt{c5.large} instance and find that it only incurs approximately a $2-3\%$
overhead on average over having no protection.

\fakesec{Architecture-dependent Optimization.} Inspired by 
LowFat~\cite{Duck2016:LowFatSW}, we use the Bit Field Extract \texttt{BEXTR} instruction 
to enhance performance. This \texttt{x86\_64} instruction
extracts~$n$ contiguous bits from a given source. \texttt{BEXTR} allows us to save
a few instructions when operating on permutations packed into integers. Instead
of shifting and applying a mask to extract the necessary bits corresponding to a
permutation (e.g., Block Offset bits in~\cref{fig:perm-runtime}),
we replace this with a single instruction.

\section{Evaluation}\label{sec:evaluation}

We evaluate~\pname{} across multiple dimensions. First, we compare the 
performance of~\pname{} against state-of-the-art pre- and post-deployment memory 
safety solutions using SPEC2017. Second, we demonstrate~\pname{}'s deployability by 
compiling and running three real-world applications. Third, we 
analyze~\b2p{}'s completeness by reporting its coverage for all benchmarks. 
Fourth, we evaluate~\pname{}'s suitability for 
multi-threaded applications. Fifth, we measure the uniformity and efficiency 
of~\pname{} permutations to support our security claims. 

\fakesec{Experimental Setup.} We run our experiments on a bare-metal Intel Skylake-based
Xeon Gold 6126 processor running at 2.6GHz with RHEL Linux 7.5 (kernel 3.10). We
compare~\pname{} against AddressSanitizer
(ASAN) and Intel MPX, as representatives of pre- and post-deployment memory 
safety solutions, respectively.\footnote{We exclude SoftboundCETS 
  as it fails to run many of the benchmarks due to strictness and 
  compatibility issues.} Each tool is
run using its best recommended settings (See~\cref{tab:compiler-setup} for a full
list of compiler flags and environment variables). We run each tool
such that it suppresses its warnings or errors so that benchmarks run to
completion. Additionally, we disable any reporting to minimize the performance
impact this functionality may have. For tools that provide
coverage other than the heap, we only enable their respective heap support for a
fair comparison with our current~\pname{} prototype.\footnote{While our current 
prototype supports stack memory, there are still minor items that have yet to be 
completed to make it robust enough for all programs. Thus, we leave the  
evaluation of stack memory permutation for future work.} Given the difference in
compiler versions and optimization levels that each tool supports, we normalize
each against their respective baselines for proper comparison. To minimize
variability, each benchmark is executed 5 times and the average of the execution
times is reported; error bars represent
the maximum and minimum values observed over the 5 runs.

\begin{table}[!t]
  \centering
  \caption{Compiler Setup}\label{tab:compiler-setup}
  \vspace{1em}
  \resizebox{0.48\textwidth}{!}{%
  \begin{tabular}{@{}llll@{}}
  \toprule
  \textbf{Tool}           & \textbf{Compiler} & \textbf{Compile Flags}                                                                                      & \textbf{Runtime Flags}                                                                                                                     \\ \midrule
  ASAN & Clang 9           & \begin{tabular}[c]{@{}l@{}}\texttt{-fsanitize=address}\\ \texttt{-asan-stack=0} \\ \texttt{-asan-globals=0}\\ \texttt{-fsanitize-recover=address}\\ \texttt{-O3}\\ \texttt{-flto=thin}\end{tabular}                                    & \begin{tabular}[c]{@{}l@{}}\texttt{print\_legend=false}\\ \texttt{print\_full\_thread\_history=false}\\ \texttt{halt\_on\_error=false}\\ \texttt{detect\_leaks=0}\end{tabular} \\ \midrule
  MPX               & GCC 7.3.1         & \begin{tabular}[c]{@{}l@{}}\texttt{-fcheck-pointer-bounds}\\ \texttt{-fchkp-narrow-bounds}\\ \texttt{-fchkp-use-wrappers}\\ \texttt{-O2}\end{tabular} & \begin{tabular}[c]{@{}l@{}}\texttt{CHKP\_VERBOSE=0}\\ \texttt{CHKP\_PRINT\_SUMMARY=0}\end{tabular}                                                           \\ \midrule
    \begin{tabular}[c]{@{}l@{}}\pname{}\\(\textit{This paper})\end{tabular}             & Clang 4.0.0       & \begin{tabular}[c]{@{}l@{}}\texttt{--size-128}\\ \texttt{--func-scope-opt}\\ \texttt{--cache}\\ \texttt{-O3} \\ \texttt{-flto=thin}\end{tabular}                           &                                                                                                                                            \\ \bottomrule
  \end{tabular}%
  }
\end{table}

\subsection{Performance Comparisons}

\begin{figure}[!h]
  \centering
  \includegraphics[width=0.49\textwidth]{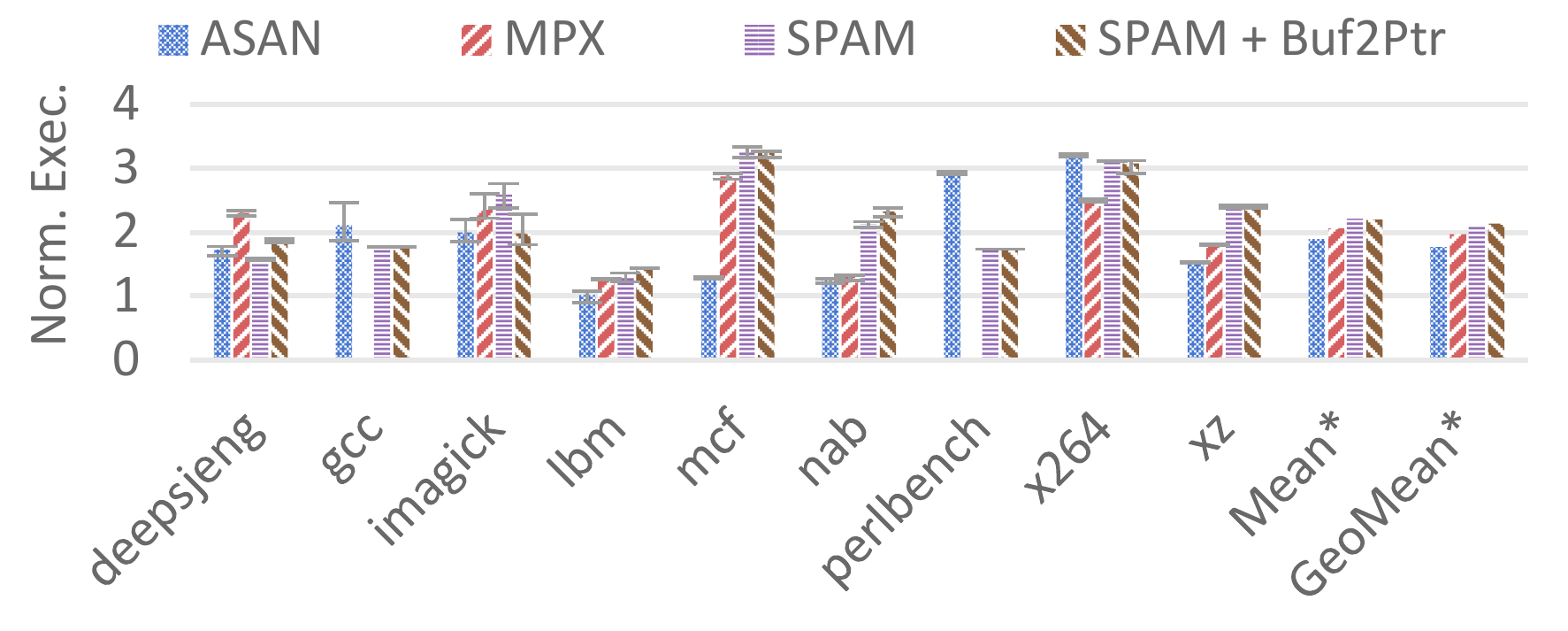}
  \caption{Performance overheads of the C subset of SPEC2017 for different tools normalized to
    their corresponding baseline for heap memory protection.}\label{fig:spec-perf}
\end{figure}

We compare~\pname{}'s performance against ASAN and Intel MPX by compiling 
and running a standard application benchmark suite, namely the C
programs in SPEC2017~\cite{Bucek2018:SPEC}, using the three tools separately. 
Additional benchmarks (\eg{}
Olden~\cite{Anne1995:Olden} and PtrDist~\cite{Todd1995:PtrDist}) are
evaluated in \cref{sec:eval-appendix}. While analyzing the results, it is
important to keep in mind the security coverage that each tool provides because they
are not all the same.

\fakesec{SPEC2017.} We specifically look at the C subset of benchmarks in SPEC
CPU2017~\cite{Bucek2018:SPEC}. Of the 9 C benchmarks, we found that \texttt{perlbench}
and \texttt{gcc} are heavily reliant on undefined behavior, namely using out-of-bounds pointers 
in computations, storing them in memory, and returning them in-bounds 
again upon pointer dereferencing. The undefined
behavior in \texttt{perlbench} and \texttt{gcc} makes them incompatible with
\pname{} and MPX. For the purposes of evaluating \pname{}'s performance, we run
the two troublesome programs with an
``in-order'' permutation (i.e., all of the~\pname{}'s instrumentation and runtime wrappers 
are used whereas the program data itself is written to memory with no shuffling) to closely 
model \pname{}'s impact. We omit
\texttt{perlbench} and \texttt{gcc} from the computed averages in the figure.
Benchmarks are run
to completion using the \texttt{test} inputs and single threaded execution. For
benchmarks with multiple inputs, the sum of the execution time of all inputs is
used. The geometric mean of each tool is as follows: ASAN~($1.77$x), 
MPX~($1.97$x), \pname{}~($2.11$x)\footnote{Enabling global memory permutation 
for \pname{} incurs an additional $11\%$ on top of the heap results bringing the geometric mean
from $2.11$x to $2.22$x.},
and \pname{} + \b2p{} ($2.13$x). The main reason for the high overheads of certain
SPEC2017 benchmarks, such as \texttt{mcf} and \texttt{x264} is the excessive use
of \texttt{memcpy}. Although our current prototype implements its own
\texttt{memcpy} wrapper for performance, it is not as efficient as the standard library one with
vectorization support. We leave the implementation of more efficient library wrappers for future work.

\begin{figure}[!h]
  \centering
  \includegraphics[width=0.35\textwidth]{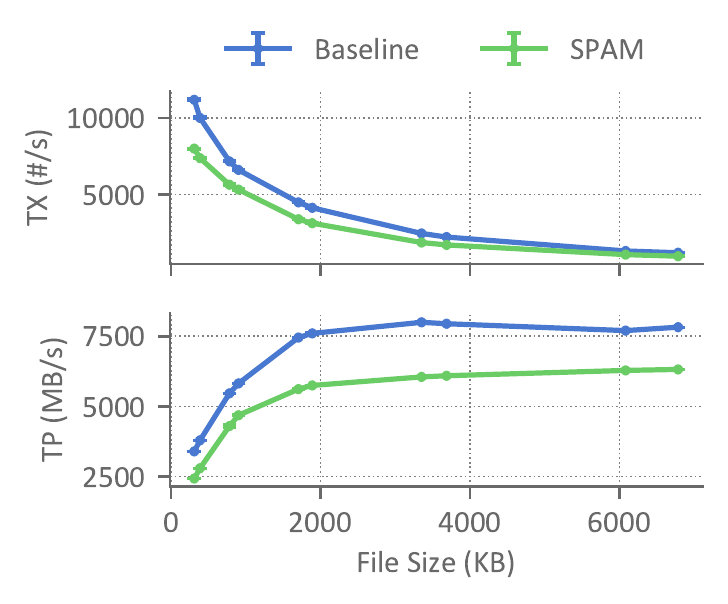}
  \caption{Nginx performance overheads with \pname{} for heap memory protection.}\label{fig:nginx-perf}
\end{figure}

\subsection{Real-world Case Studies}

To demonstrate the capabilities of our current~\pname{} prototype, we 
use it to compile and run three real-world applications: the Nginx web 
server~\cite{nginx}, the Duktape Javascript interpreter~\cite{duktape}, 
and the WolfSSL cryptographic library~\cite{wolfssl}.

\fakesec{Nginx.} We use Nginx~\cite{nginx} (version $1.11.11$),
as a representative I/O bound benchmark.
To simulate typical workload configurations, we have Nginx serve different sized
files according to the page weight (\ie{} the amount of data served) of modern
websites according to the 2019 HTTP Archive Web Almanac
report~\cite{httparchive}. To generate the client load, we used the multi-threaded
Siege~\cite{siege} benchmarking tool. We issued $500$ requests with $50$
concurrent connections for each page weight using the loopback interface to avoid network congestion
issues. We
record the throughput (TP), and transfer rates (TX). The results
in~\cref{fig:nginx-perf} show that on average \pname{} incurs a $1.3$x overhead
relative to the baseline. As file sizes become larger, the I/O starts to
dominate and the performance impact of \pname{} is as low as $1.24$x; in
contrast, for smaller files \pname{}'s performance impact increases to $1.4$x.
Enabling global memory permutation shows no measurable performance difference over the heap
results. This is primarily due to the fact that the initial cost of permutation
for globals (\ie{} \runtimefn{RegisterGlobal}) is amortized as the server is
already loaded before it receives requests.

\begin{figure}[!h]
  \centering
  \includegraphics[width=0.35\textwidth]{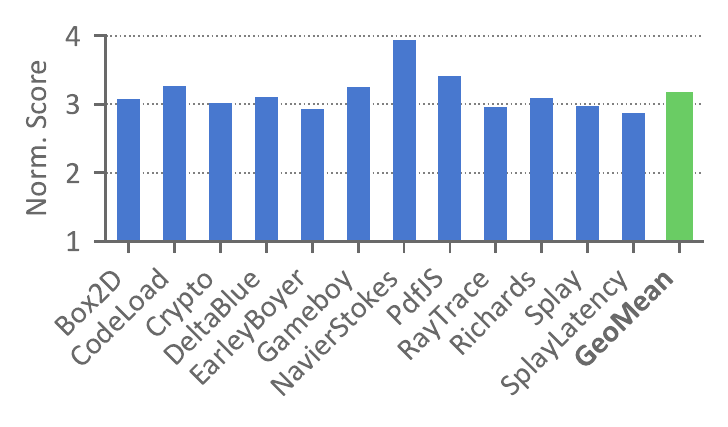}
  \caption{Duktape performance overheads with \pname{} for heap memory protection.}\label{fig:duktape-perf}
\end{figure}

\fakesec{Duktape.} We evaluate the Duktape~\cite{duktape} (version $2.5.0$)
Javascript interpreter with a default build configuration running the Octane 2 benchmark
suite~\cite{octane2}\footnote{Not all benchmarks are supported by Duktape.}.
We record the benchmark scores as reported by the Octane 2 suite and show the relative
performance of \pname{} compared to the baseline. The results
in~\cref{fig:duktape-perf} show an overhead with a geometric mean of
approximately $3.15$x. Similar to Nginx, enabling global memory permutation shows no measurable performance
overheads for much the same reasons. The interpreter instance is already loaded by the time the Octane 2 benchmarks begin, thus amortizing the permutations associated with \runtimefn{RegisterGlobal}.

\begin{figure}[!h]
  \centering
  \includegraphics[width=0.49\textwidth]{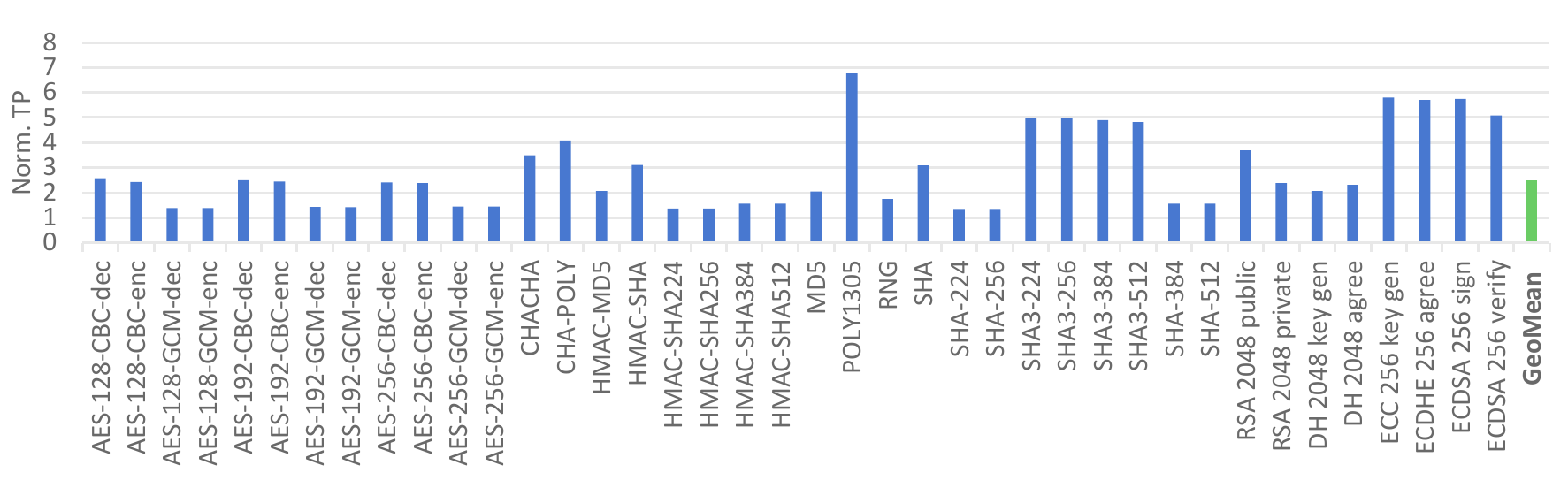}
  \caption{WolfSSL performance overheads with \pname{} for heap memory protection.}\label{fig:wolfssl-perf}
\end{figure}

\fakesec{WolfSSL.} We evaluate WolfSSL~\cite{wolfssl} (version $4.4.0$), as it is a
popular cryptographic library. We use the default build configuration and the
included wolfCrypt Benchmarks with default parameters which measure symmetric
algorithms, such as AES and ChaCha20 and asymmetric algorithms, such as RSA and ECC in
terms of throughput. The results in~\cref{fig:wolfssl-perf} are normalized
against a baseline execution and show an overhead with a geometric mean of
$2.48$x. Enabling global memory permutation increases the average overheads by an additional
$16\%$.

\subsection{\b2p{} Analysis}\label{subsec:b2p-analysis}

\begin{figure}[!t]
  \centering
  \includegraphics[width=0.4\textwidth]{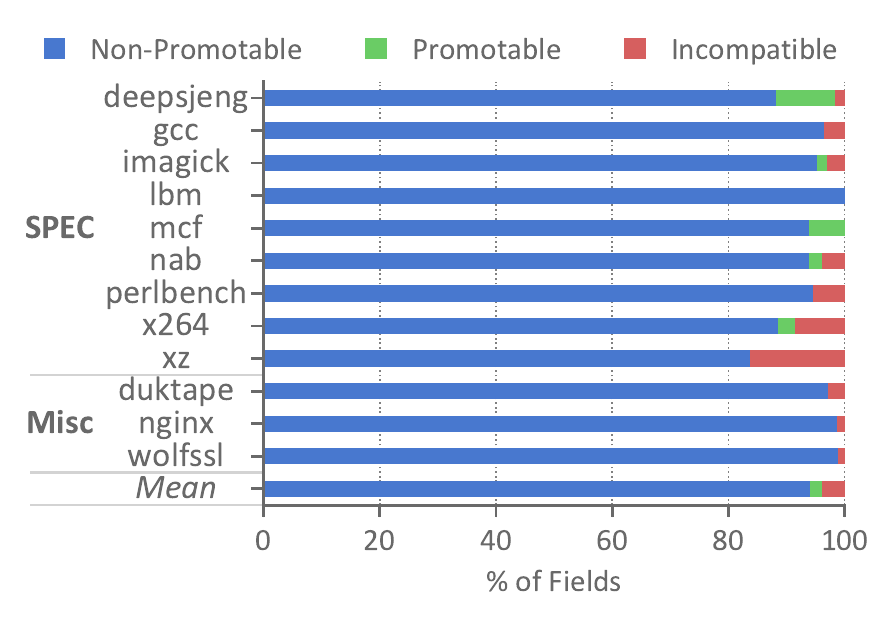}
  \caption{Percentage of total struct fields covered by the Buf2Ptr transformation.}\label{fig:buf2ptr-coverage}
\end{figure}

\fakesec{Coverage.} We show the coverage
of our current \b2p{} implementation in~\cref{fig:buf2ptr-coverage}.
Coverage is reported in terms of the total number of fields in
an application with each field belonging to one of three categories: (1)
\textit{Non-promotable}, means that \b2p{} is unnecessary as there are no array
fields in an object (2) \textit{Promotable},
means an array field which can be safely promoted and (3) \textit{Incompatible}, means an array
field that is currently considered unsafe to promote. The data shows that the majority of
fields are considered to be non-promotable~($94\%$ on average). Defining structures with internal
array fields is fairly uncommon among the benchmarks we sample~($6\%$ on average). As a result, the
total number of promoted fields is quite low.

The fields that are considered to be incompatible can be grouped into two
categories: (1) those that are fundamentally troublesome to the \b2p{} approach (see
\cref{lst:b2p-considerations} for detailed examples) and (2) artifacts of
our current implementation. The ones that result from our current implementation
artificially inflate our results for incompatible fields. The most note-worthy
is the limited support for promoting fields wrapped in macros. As it stands, the
current Clang rewriter API has limited support for macro rewriting. An
alternative to this current limitation is to expand macros before processing in
the frontend, but this comes with its own drawbacks making it
difficult to preserve the syntactic structure of the program. For our current
implementation, we have chosen to forgo this route in favor of preserving the
definition and use of macros and other formatting in the source code. This
choice primarily affects the \texttt{imagick} and \texttt{x264} benchmarks which
use a \texttt{CHECKED\_MALLOC} macro for all structs. In addition to macros, our
current implementation has limited support for embedded struct declarations,
variable length arrays and single statement variable declarations. These
are straightforward extensions and are left for future work.

\fakesec{Discussion.} The (\pname{} + \b2p{}) column in~\cref{fig:spec-perf} shows the additional performance overheads of applying
\b2p{} compared to \pname{} alone. We notice that the
performance overheads in \b2p{} are mainly due to (1) the one extra pointer access
operation per each sub-object array and (2) the poor locality of sub-object arrays as
they are allocated in different regions. The cost of the extra pointer operation can
be easily amortized if there exist multiple successive accesses to the same
sub-object array with different addresses. In this case, one possible
optimization is to only compute the new base address of the sub-object once and
allow all future accesses to use it.
The locality issue highly depends on the application behavior. If the
application tends to access all struct fields one after the other, locality and the processor's
cache hit rate will be affected. If the application tends to access the
same field from different objects, locality will not
matter.

From a security perspective, the low number of promotable fields we have
observed in the evaluated programs has a few interesting implications.
First, it suggests that \emph{intra-object overflows may account for a small
portion of the attack surface in user-space programs.} However, it does not help us quantify the
severity of intra-object overflow vulnerabilities. Second, with respect to
\b2p{}, the low number of promotable fields means that the scope of
incompatibilities is limited. Thus, one option to addressing the \b2p{} limitations
discussed in \cref{sec:limitations} may be done with minimal developer effort
(\eg{} annotations).

\subsection{Multi-Threaded Applications}

\begin{figure}
  \centering
  \includegraphics[width=0.45\textwidth]{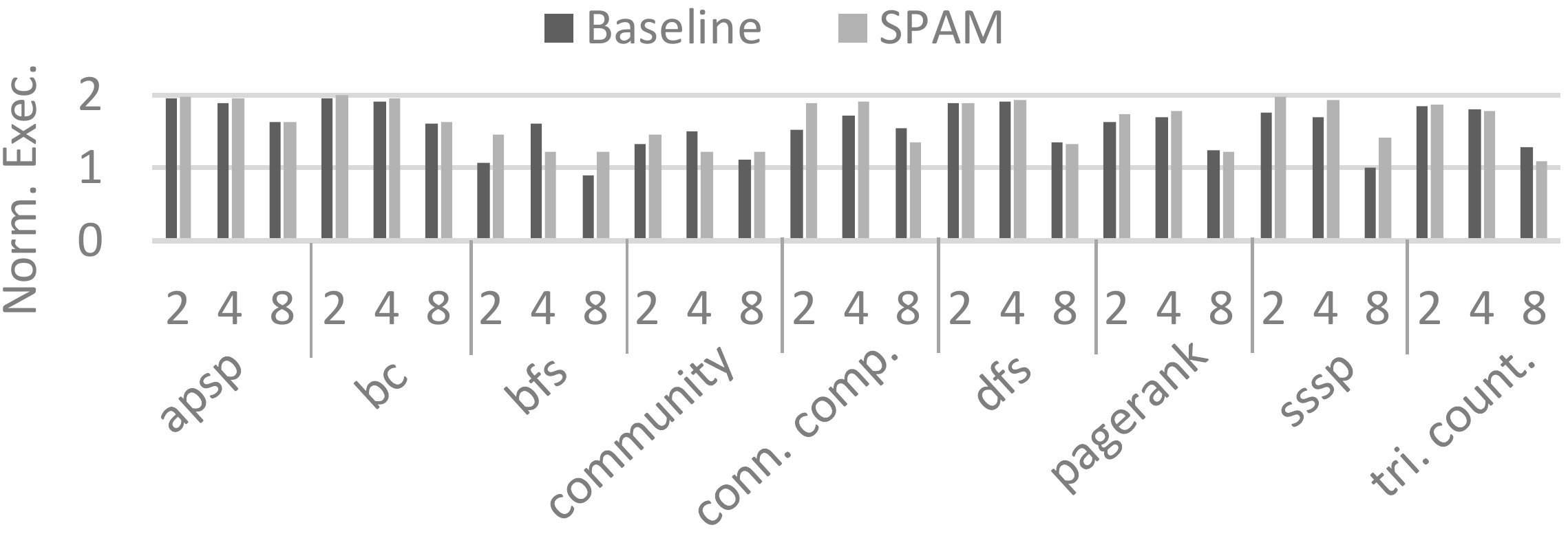}
  \caption{\pname{}'s multi-threaded scalability 
    for 2, 4, and 8 threads relative to the single threaded program.}\label{fig:crono-multi}
\end{figure}

State-of-the-art memory safety techniques (e.g., ASAN, Intel MPX)
maintain explicit metadata for pointers and objects. This can result in false
positives/negatives in multi-threaded programs if the metadata is not atomically
updated in the same transaction as a program's atomic updates to its associated
pointers or objects. The stateless nature of \pname{} makes it well suited for
multi-threaded applications. To verify this hypothesis, we run the CRONO benchmark
suite~\cite{Ahmad2015:CRONO}, a specialized multi-threaded suite of graph
algorithms. Through successful instrumentation and execution of benchmark
applications, it is shown that our proposed solution is thread-safe and is
suitable for multicore systems. We perform a sweep of each benchmark
using~$2$,~$4$, and~$8$ threads with each bar showing the normalized execution relative to
the single threaded program. The baseline bars show how the performance scales
for the default program with the \pname{} bars representing the scaling for
the hardened version. The results in~\cref{fig:crono-multi} show that the relative
performance improvements of multi-threaded programs are unaffected by our
stateless runtime as the bars for the baseline and \pname{} are almost equivalent.

\subsection{Permutation Analysis} \label{subsec:rand-eval}

\begin{figure*}[ht!]
    \centering
    \begin{subfigure}[t]{0.32\textwidth}
        \centering
        \includegraphics[width=\textwidth]{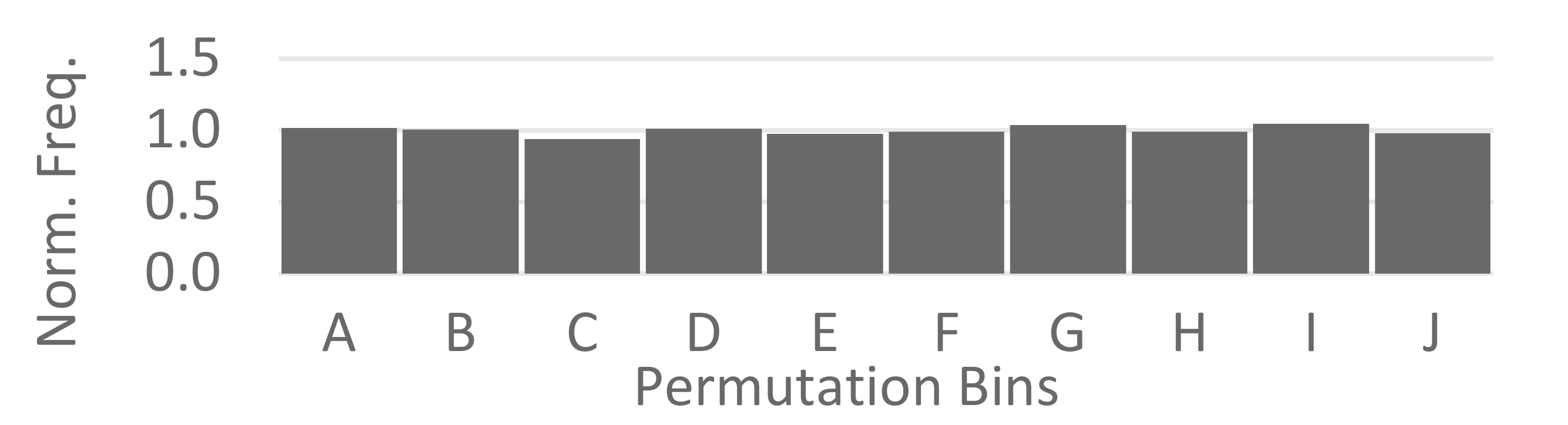}
        \caption{Intra-Allocation Region}\label{fig:perm-intra}
    \end{subfigure}%
    ~ 
    \begin{subfigure}[t]{0.32\textwidth}
        \centering
        \includegraphics[width=\textwidth]{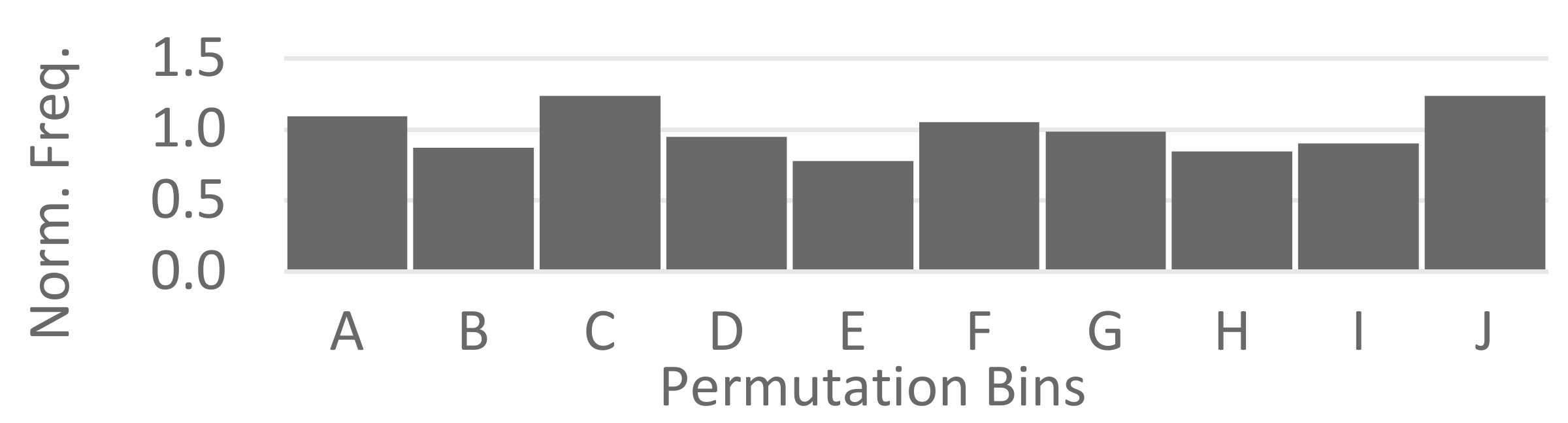}
        \caption{Inter-Allocation Region}\label{fig:perm-inter}
    \end{subfigure}
    ~ 
    \begin{subfigure}[t]{0.32\textwidth}
        \centering
        \includegraphics[width=\textwidth]{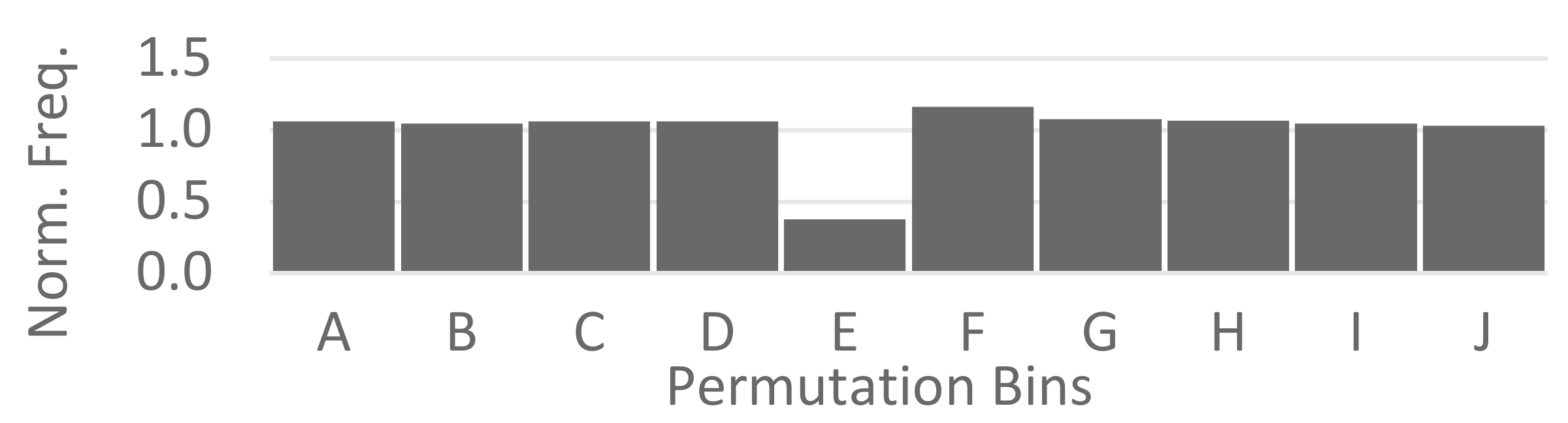}
        \caption{Alias Number}\label{fig:perm-alias}
    \end{subfigure}
    \caption{The distribution of generated permutations (a) within the same allocation
    region (b) across different allocation regions and (c) for aliases of the
    same allocation.}
\end{figure*}

\fakesec{Uniformity.} As discussed
in \cref{sec:securityanalysis}, the uniformity of permutations is important 
as it reduces the success probability of an attacker.
To evaluate the uniformity of permutations within our
framework we conducted three
experiments: (1) within (\textit{intra}) allocation regions, (2) across
(\textit{inter}) regions, and (3) different alias numbers for the same address.
The tests are designed to stress the permutation scheme. We use allocation
regions of multiples of 128B
for our LowFat allocator. For the intra region experiment
(\cref{fig:perm-intra}), we perform one million allocations of a given size, in
this instance the smallest region (128B). For the inter region experiment
(\cref{fig:perm-inter}), we perform 100 allocations for 20 different LowFat
regions (\ie{} 2000 total allocations). For the aliasing experiment
(\cref{fig:perm-alias}), we perform 10,000 allocations using the same address
(so that only alias number bits are different).
For all experiments, we plot histograms with 10 bins. The values are normalized
according to the expected value of each bin (\eg{} the total number of allocations
divided by total number of bins). As the data
shows, the results in ~\cref{fig:perm-intra,fig:perm-inter} are very close to
ideal indicating that any permutation is equally likely of being generated.
\cref{fig:perm-inter} has slightly higher variability compared
to~\cref{fig:perm-intra} simply due to the smaller number of allocations. 
\cref{fig:perm-alias} shows slight modulo biasing as Fisher-Yates must bound the
range of the values from the PRNG, as discussed in \cref{subsec:spatial}. 
While it is theoretically possible to remove
this bias, it involves the possibility of indefinitely polling the PRNG. We
leave the exploration of alternative FPE shuffling
schemes~\cite{DBLP:conf/crypto/HoangMR12,DBLP:conf/crypto/RistenpartY13} to
future work.

\begin{table}[!h]
  \centering
  \caption{Throughput of generating permutations for various PRNGs tested.}\label{tab:perm-throughput}
    \footnotesize
    \begin{tabular}{ll}
      \toprule
      \textbf{PRNG} & \textbf{Cycles} \\ \midrule
      AESRand       & 504             \\
      xorshift64*   & 870             \\
      xorshift128+  & 932             \\
      rand          & 4652            \\ \bottomrule
    \end{tabular}%
\end{table}

\fakesec{Performance.} A cornerstone in our framework is the ability to generate truly random
permutations. We tested a handful of pseudo-random number generators (PRNGs)
available~\cite{xorshift,rand,AESRand}. \cref{tab:perm-throughput} shows the throughput of generating permutations for
the various PRNGs ultimately leading to our choice of AESRand~\cite{AESRand} as it was the most
performant within our framework. This fact is critical in keeping performance
overheads low as the PRNG lies within the critical path of our entire approach.
AESRand is based on the Intel AES-NI instruction set extensions and 128-bit
SIMD. It is known to pass two widely used statistical tests for PRNGs, namely
Big Crush~\cite{BigCrush} and PractRand~\cite{PractRand}.

\subsection{Summary}\label{subsec:summary}
Unsurprisingly, our evaluation shows that \pname{} suffers from high performance
overheads. This is expected due to (1) the additional instrumentation instructions
added for every load and store; and most importantly, (2) the software implementation of
the permutation function. Both of these overheads can be mitigated with simple
hardware support. For example, the \texttt{getaddr} primitive can be combined with a load
or store in a single new instruction reducing code size and enhancing
instruction cache utility. Additionally, the \texttt{genperm} primitive can be implemented
with a hardware permutation network that reduces the current permutation latency
from~$504$ cycles to be within~$2$ cycles~\cite{Lee2001:permute,Shi2003:two-cycle}.
We leave the design space exploration of \pname{}'s hardware support for future
work. \emph{Even without hardware support, the cohesive security of \pname{}
provides stronger guarantees than existing solutions at modest performance
overheads (\eg{} Nginx).}

\section{Deployment Considerations}\label{sec:deploy}

\pname{} requires a strict separation between application code and data 
(permuted domain) and external code and data (unpermuted domain). Otherwise, the 
loads and stores in the external code may inadvertently corrupt application 
memory as the external code (unless compiled with \pname{}) is not instrumented 
to deal with permuted memory. In this section, we discuss different cases that 
affect the boundaries and how we handle them. 

\fakesec{Externally Invoked Function Pointers.} Instrumented code may be called
externally via a function pointer. It is possible that the function is 
called from a non permuted domain. One example, is the comparison function
pointer argument in \incode{qsort}. This comparison function is typically
defined by the application and thus will be instrumented. On the other hand,
\incode{qsort} itself is not instrumented as it belongs to external libraries.
Depending on the logic of the external call that uses the function pointer we
have the option of choosing to either (1) instrument the external library and
all of its dependencies so that they are part of the permuted domain, or (2)
create a variant of the function being pointed to by the function pointer and
all of its dependencies that recursively unpermute the data upon load. Of the
benchmarks we evaluate in Section~\ref{sec:evaluation}, we encountered this scenario
with \incode{qsort} in \texttt{imagick,nab} and \texttt{x264} and handle them
using the second option.

\fakesec{External Calls \& Nested Memory.} Depending on the API of the used external
library, there may be instances in which pointers to permuted memory may be
stored inside nested structures. \pname{} handles these nested cases by relying
on the type information of the function arguments to recursively unpermute and
permute memory as necessary. The main limitation with the recursive approach is
with self referencing structures (\ie{} those with cycles). In this case, we
support specifying a recursion limit for a given function. Alternatively, a
suitable \pname{} wrapper can be implemented to optimally handle 
the recursion depth necessary for a given function.

\fakesec{External Calls \& Multi-threading.} For multi-threaded programs, the
semantics of the external call determine the instrumentation that our compiler
emits. For external functions that only read permuted memory (\eg{}
\texttt{memcpy}), we emit a single copy unpermute primitive avoiding contention
between other threads who may be trying to read the same memory while being
unpermuted. If the external function tries to write to permuted memory, this is
a race condition in the original code and should be resolved by the original
developer using a synchronization primitive. In this case, \pname{} emits the
regular unpermute primitive.

\fakesec{Externally Allocated Memory.} A number of functions in \texttt{libc}
may return dynamically allocated memory, the majority of these are lumped under
the \textit{Dynamic Allocation Functions}~\cite{ISO24731} extension (\eg{}
\incode{strdup}, \incode{getline}, etc). While nothing is required for
correctness to use these functions with \pname{}, memory returned by them would
be unpermuted. In order to protect memory in these situations, our prototype emits 
a permute primitive for memory returned by these functions.
\section{Limitations}\label{sec:limitations}

In this section, we discuss the limitations of our current implementation and how we 
plan to address them in the future.

\begin{listing}[!h]
  \centering
    \begin{minipage}{0.44\textwidth}
      \begin{minted}
        [
        linenos,
        fontsize=\footnotesize,
        mathescape,
        autogobble,
        stripnl=false,
        numbersep=3pt
        ]
        {cpp}
        struct Foo *f = malloc(sizeof(struct Foo));

        // Type Erasure $\label{line:b2p-type-erasure}$
        void *type_erasure = (void *) f; 
        // Type Confusion $\label{line:b2p-type-confusion}$
        struct Bar *type_confusion = (struct Bar *)f;

        // Anonymous Allocation $\label{line:b2p-anonymous}$
        void *anon_alloc = malloc(10);
        f = anon_alloc;
        struct Foo *f2 = anon_alloc + sizeof(struct Foo);

        // Struct Return-by-Value $\label{line:b2p-ret}$
        struct Foo getCopy(struct Foo f);
      \end{minted}
    \end{minipage}
  \caption{Examples of code that cannot be legally transformed using our current~\b2p{} prototype.}\label{lst:b2p-considerations}
\end{listing}

\fakesec{\b2p{}.}
\cref{lst:b2p-considerations} shows the different types of legality constraints
that apply to \b2p{}. For example, type erasure and type confusion result in the loss
of the original type information due to implicit or explicit casting (\eg{}
Line~\ref{line:b2p-type-erasure} \& \ref{line:b2p-type-confusion}). With the aid
of alias analysis, it is possible to recover the original type in order to
safely apply the transformation. A second legality requirement pertains to what
we refer to as anonymous allocations (\eg{} Line~\ref{line:b2p-anonymous}), or
allocations in which the type is unknown. This pattern is common in applications
with custom memory allocation wrappers that manually manage large memory chunks.
In this situation, \b2p{} relies on the developer to provide type information
for a safe promotion. Finally, while passing a structure by value is supported,
returning it by value (\eg{} Line~\ref{line:b2p-ret}) is not amiable, to
promotion as it breaks the semantics of our transformation. We empirically
evaluate how often these constraints prevent us from applying \b2p{}
transformations and thus providing intra-object safety
in Section~\ref{subsec:b2p-analysis}.

\fakesec{Inline Assembly.} With the current compiler infrastructure, \pname{} is not able to
properly instrument inline assembly code. However, in the future we anticipate that with
the integration of a binary lifter~\cite{mcsema,Yadavalli2019:mctoll} we would
be able to instrument IR generated from inline assembly.

\fakesec{Variadic Functions.} Using functions with variable number of arguments 
is fully supported in our current prototype. That includes invoking variadic 
functions in instrumented code (e.g., \texttt{printf} and \texttt{scanf}). The 
only exception is invoking functions, which (1) are defined in uninstrumented 
code and (2) use \texttt{va\_list} as an argument (e.g., \texttt{vsprintf}). 
Those function invocations are not currently supported by our prototype. 
However, \texttt{va\_list} usage inside of instrumented code is fully 
supported. Support for passing \texttt{va\_list} externally is left for future 
work.

\fakesec{Additional Language Support.} Our prototype currently supports C
programs as evaluated in Section~\ref{sec:evaluation}. Introducing support for other
programming languages (e.g., C++) is straightforward as long as a clear
separation between instrumented and uninstrumented code is established. This
avoids unnecessary unpermute/permute overheads. In the case of C++, one way to
add support would involve compiling the standard C++ library (\eg{}
\texttt{libc++} or \texttt{libstdc++}) with \pname{} so that data is unpermuted
only at the system call interface.
\section{Related Work}\label{sec:relatedwork}

\begin{savenotes}
\begin{table*}[t]
  \centering
  \caption{Comparison with prior works.}\label{tab:comparison}
  \resizebox{1.0\textwidth}{!}{
  \begin{threeparttable}
  \begin{tabular}{r c c c c c c c c c l}
    \toprule

    \multicolumn{1}{c}{\multirow{2}{*}{\textbf{Proposal}} } & \multicolumn{1}{c}{\textbf{Commodity}} & \multicolumn{1}{c}{\textbf{Failure}}          & \multicolumn{1}{c}{\textbf{Deployment}}  & \multicolumn{1}{c}{\textbf{Instrumentation}} & \multicolumn{2}{c}{\textbf{Spatial Protection}}                          & \multicolumn{1}{c}{\textbf{Temporal}}   & \multicolumn{1}{c}{\textbf{Multi-threaded\tnote}} &  \multicolumn{1}{c}{\textbf{Side-channels}}   & \multicolumn{1}{c}{\multirow{2}{*}{\textbf{Main Limitations}} }  \\
                                                            & \multicolumn{1}{c}{\textbf{Systems}}   & \multicolumn{1}{c}{\textbf{Model~\tnote{*}}}  & \multicolumn{1}{c}{\textbf{Stage}} & \multicolumn{1}{c}{\textbf{Level}} & \multicolumn{1}{c}{\textbf{Inter}} & \multicolumn{1}{c}{\textbf{Intra}}  & \multicolumn{1}{c}{\textbf{Protection}} & \multicolumn{1}{c}{\textbf{Support~\tnote{$\P$}}}   &  \multicolumn{1}{c}{\textbf{Resiliency~\tnote{$\S$}}}      &   \\
    \midrule
  AddressSanitizer~\cite{Serebryany2012:ASAN}       & \cmark & Precise            & Pre    & Source    & \cmark & \xmark\tnote{$\|$} & \cmark  & \pie{180} & \pie{0}               & Detect subset of intra-object violations \\
  EffectiveSAN~\cite{Duck2018:effectiveSAN}         & \cmark & Precise            & Pre    & Source    & \cmark & \cmark & \cmark\tnote{$\dagger$}  & \pie{180} & \pie{0}               & Metadata vulnerable to memory disclosure attacks  \\
	CUP~\cite{Burow2018:CUP}                        & \cmark & Precise            & Pre    & Source    & \cmark & \xmark & \cmark  & \pie{0}   & \pie{0}               & Metadata vulnerable to memory disclosure attacks \\  
  \midrule
  SoftBoundCETS~\cite{Nagarakatte2009:SoftBound,
    Nagarakatte2010:CETS}                           & \cmark & Precise            & Post    & Source    & \cmark & \cmark & \cmark  & \pie{0}   & \pie{0}               & No support for multithreading \\
    DFI~\cite{DBLP:conf/osdi/CastroCH06} & \cmark & Precise & Post & Source & \cmark & \cmark\tnote{$\ddagger$} & \xmark & \pie{180} & \pie{0} & Imprecise points-to analysis\\
      \midrule
    Isomeron~\cite{isomeron2015} & \cmark & Imprecise & Post & Binary & \xmark & \xmark & \xmark & \pie{360} & \pie{0} &  No data protection\\
    Shuffler~\cite{David2016:shuffler} & \cmark & Imprecise & Post & Binary & \xmark & \xmark & \xmark & \pie{360} & \pie{0} &  No data protection\\
  \midrule
    MPX~\cite{oleksenko2018intel,Zhang2019:BOGO}      & \cmark & Precise            & Post   & Source    & \cmark & \cmark & \cmark  & \pie{180} & \pie{180}               & Dropped support in GCC/Linux \\
    Intel CET~\cite{IntelCeT} & \xmark & Precise & Post & Source & \xmark & \xmark & \xmark & \pie{360} & \pie{0} & No data protection\\
  ARM PAC~\cite{Qualcomm2017}                       & \cmark & Precise            & Post   & Source    & \cmark & \xmark & \cmark  & \pie{360} & \pie{180}               & Pointers-only protection \\
  ARM MTE~\cite{ARM2019:MTE}                        & \xmark & Precise            & Post   & Source    & \cmark & \xmark & \cmark  & \pie{360} & \pie{180}               & Limited entropy (4-bits) \\
  CHERI~\cite{Watson2015:CHERI, 
    Hongyan2019:CHERIvoke}                          & \xmark & Precise            & Post   & Source    & \cmark & \xmark & \cmark  & \pie{360}    & \pie{180}               & No support for intra-object protection \\
  Califorms~\cite{Sasaki2019:Califorms}             & \xmark & Precise            & Post   & Source    & \cmark & \cmark & \cmark  & \pie{360}    & \pie{180}               & No runtime randomization \\
  \midrule
  PointGuard~\cite{Crispin2003:Pointguard}          & \cmark & Imprecise          & Post   & Source      & \cmark & \xmark & \xmark  & \pie{360} & \pie{0}               & Pointers-only protection. Weak encryption (XOR)  \\
  Data Randomization~\cite{Costa2008:MicrosoftDR, 
     Bhatkar2008:DSR}                               & \cmark & Imprecise          & Post   & Source    & \cmark & \xmark & \xmark  & \pie{360} & \pie{360}               & Same layout for all instances. Weak encryption (XOR) \\
  SALADSPlus~\cite{Chen2015:salads, 
     Chen2018:saladsPlus}                           & \cmark & Precise            & Post   & Source      & \cmark & \cmark & \xmark  & \pie{180} & \pie{180}               & Partial objects randomization \\
    Shapeshifter~\cite{Wang2020:Shapeshifter}       & \cmark & Precise            & Post   & Source     & \cmark & \cmark & \xmark  & \pie{180} & \pie{180}               & Metadata vulnerable to memory disclosure attacks  \\
  POLaR~\cite{Kim2019:Polar}                        & \cmark & Imprecise          & Post   & Source      & \cmark & \cmark & \xmark  & \pie{180} & \pie{180}               & Partial objects randomization \\
  Smokestack~\cite{Aga2019:Smokestack}              & \cmark & Imprecise          & Post   & Source      & \cmark & \xmark & \xmark  & \pie{180} & \pie{0}               & Stack-only protection \\
  RA-malloc~\cite{Jang2019:RUMA}                    & \cmark & Imprecise          & Post   & Binary   & \cmark & \xmark & \xmark  & \pie{360} & \pie{0}               & Low randomization entropy (3-bits) \\
  \midrule
  \textbf{\pname{}}                                 & \cmark & Imprecise          & Post   & Source    & \cmark & \cmark & \cmark  & \pie{360} & \pie{360}               & Incomplete type information in \b2p \\
  \bottomrule
  \end{tabular}
  \vspace{1em}
  \begin{tablenotes}
    \item [*] Solutions with imprecise failure models are
      suitable for post-deployment. Those with precise failure models are
      suitable for pre-deployment (\ie{} testing).
    \item [$\P$] \pie{360} - Supported (stateless); \pie{180} - Supported
      (requires synchronization on global metadata); \pie{0} - No support. 
    \item [$\S$] \pie{360} - Resilient to RowHammer, hardware side-channels,
      ColdBoot; \pie{180} - Resilient to some; \pie{0} - None.
    \item[$\|$] Experimental support via \texttt{-fsanitize-address-field-padding}
    \item[$\dagger$] Does not detect errors where the free'd object is reallocated to an object of the same type.
    \item[$\ddagger$] Depends on points-to analysis.
  \end{tablenotes}
  \end{threeparttable}
  }
\end{table*}
\end{savenotes}

~\cref{tab:comparison} summarizes how~\pname{} compares to prior work. 
For each proposal, we specify its security 
guarantees, instrumentation level, and main limitations. 
We also highlight whether the proposal is available 
on commodity systems, has a precise failure model, supports multi-threading, 
and provides side-channel resiliency or not. Finally, we specify whether the proposal is 
used as a pre-deployment testing tool (i.e., Sanitizer) or a post-deployment mitigation.

\fakesec{Software-based Memory Safety Techniques.} 
Many solutions have been proposed in academia and industry to tackle the memory 
unsafety problem of low level languages. We divide them into two categories: 
software and hardware. 

Software only solutions can address a wide range of memory
errors either pre-deployment (i.e., Sanitizers) or post-deployment (i.e., defenses). 
For instance, AddressSanitizer~\cite{Serebryany2012:ASAN} uses shadow memory to detect 
out-of-bounds errors and uninitialized reads, respectively. However, it lacks 
the intra-object protection provided by~\b2p{}. Additionally, while ASAN is great at
finding memory corruption during testing, it cannot be used as a mitigation. An 
attacker who knows that ASAN is in use can simply move pointers past red 
zones.~\pname{} avoids such limitations by using fine-grained per-instance memory 
permutation. On the other hand, EffectiveSAN~\cite{Duck2018:effectiveSAN}, a 
sanitizer that is built on top of the LowFat allocator, can detect intra-object violations. 
However, EffectiveSAN 
does not detect use-after-free errors in which the freed object is reallocated 
to an object of the same type. This case is handled by~\pname{} as every 
allocation utilizes a random alias address, regardless of its type.
Unlike~\pname{}, EffectiveSAN stores metadata separately, making 
it vulnerable to hardware memory violations, 
such as speculative execution and side-channels. Finally, CUP~\cite{Burow2018:CUP}
uses per allocation metadata to provide spatial (inter) and temporal memory
safety. This metadata is vulnerable to hardware memory corruption attacks and limits
scalability of multi-threaded applications. While CUP
instruments \texttt{libc}, compatibility with unprotected code is not supported
by default as it poses significant performance overheads. 

Software only solutions can also be used as post-deployment mitigations. For instance,  
SoftboundCETS~\cite{Nagarakatte2009:SoftBound,Nagarakatte2010:CETS} provides 
intra-object protection with a modest performance overhead. Unlike~\pname{}, 
SoftboundCETS can not support multi-threaded applications,
which highly limits its practicality. Another example is 
DFI~\cite{DBLP:conf/osdi/CastroCH06}, which utilizes
points-to analysis to enforce data-flow integrity. While it is able to provide
spatial safety it does not address temporal vulnerabilities. Moreover, DFI
uses explicit metadata which limits its scalability for multithreaded
applications. 

Other software-based exploit mitigations, such as Isomeron~\cite{isomeron2015}
and Shuffler~\cite{David2016:shuffler} focus on randomizing code layout.
Randomizing the code layout serves as an effective means to make exploits (\eg{}
ROP, JIT-ROP, etc) more difficult with minimal performance overheads. 
However, these approaches do not directly
provide protection against spatial/temporal memory safety vulnerabilities.
Additionally, as they do not modify data, they provide no resiliency to
side-channels. We categorize~\pname{} as a post-deployment mitigation that can 
reliably thwart attackers who abuse software/hardware memory vulnerabilities 
without using explicit metadata or inserting runtime checks.  

\fakesec{Hardware-based Memory Safety Techniques.} 
Hardware-assisted solutions are promising in terms of low runtime overheads. 
Industrial solutions include Intel MPX~\cite{oleksenko2018intel}, which maintains 
objects bounds in hardware registers, Intel Control-flow Enforcement 
Technology (CET)~\cite{IntelCeT}, ARM pointer authentication 
(PAC)~\cite{Hans2019:PAC} and memory tagging (MTE)~\cite{ARM2019:MTE}. 
Unfortunately, MPX introduces compilation complexities leading to 
low adoption in practice (GCC and Linux recently dropped its 
support~\cite{gccMPX, LinuxMPX}). Intel CET only provides backward-edge 
protection with a shadow stack and a coarse-grained forward edge 
protection with \texttt{ENDBRANCH} instruction. 
ARM PAC's scope is limited, only protecting code pointers such as return
addresses and function pointers. Using PAC to protect data pointers will come
with a large performance overhead. While ARM MTE provides spatial and temporal
memory safety it has a limited entropy of $1/16$ compared to \pname{}'s $1/16!$.
Moreover, MTE does not provide intra-object spatial memory safety leaving it
vulnerable to type-confusion attacks.
On the contrary,~\pname{}'s cohesive approach of permuting memory removes 
the need for concatenating incompatible defenses.

Academic proposals, such as CHERI~\cite{Watson2015:CHERI, Hongyan2019:CHERIvoke} and
Califorms~\cite{Sasaki2019:Califorms}, come with their own limitations as well. For example, 
CHERI offers no intra-object protection. Califorms~\cite{Sasaki2019:Califorms} avoids
CHERI's limitations by randomizing the layout of structs at compile time,
providing probabilistic intra-object protection. Unlike~\pname{}, Califorms' struct
randomization is exactly the same for all instances of the same struct/class
requiring that the binary remain secret. Moreover, CHERI~\cite{Hongyan2019:CHERIvoke} 
and Califorms~\cite{Sasaki2019:Califorms} provide temporal memory safety by 
using quarantining (\ie{} placing the freed memory chunks in a queue such that 
those chunk will not be returned again by \texttt{malloc} for some period of 
time). On the contrary,~\pname{} utilizes permutations, 
which allow memory chunks to be available to new allocations immediately after 
being freed, avoiding any additional performance penalty.  

\fakesec{Data Randomization.} 
Randomization stands as a last line of defense when full memory safety cannot be
completely guaranteed. We divide data randomization techniques into two categories; static and dynamic. 

Static techniques either randomize data structure layout at compile
time~\cite{Lin2009:polymorphing, Stanley2013} or randomize the representation 
of data in memory via encryption~\cite{Crispin2003:Pointguard, Bhatkar2008:DSR, 
Costa2008:MicrosoftDR}. The above techniques maintain the same layout for all 
instances of data structs, whereas~\pname{} changes the layout based on data location. 

We distinguish~\pname{} from prior efforts on dynamic data randomization as follows. 
First, SALADSPlus~\cite{Chen2015:salads, Chen2018:saladsPlus}, 
Shapeshifter~\cite{Wang2020:Shapeshifter} and POLaR~\cite{Kim2019:Polar} 
do not randomize all program data-structures to reduce runtime overheads. 
The protected subset is either manually chosen~\cite{Chen2015:salads, 
Chen2018:saladsPlus, Wang2020:Shapeshifter}, or based on data-flow 
analysis~\cite{Kim2019:Polar}. Second, the above solutions offer no protection 
for their runtime metadata. Unlike~\pname{}'s strong threat model, 
POLaR~\cite{Kim2019:Polar} assumes that the attacker has no 
read/write capabilities before bypassing the defense itself. 
Third, while Smokestack~\cite{Aga2019:Smokestack} relies on secure PRNG to pick 
runtime permutations similar to~\pname{}, it generates the permutations 
offline and stores them in lookup table at program memory. The limited size of such 
table reduces Smokestack entropy compared to~\pname{}. Additionally, Smokestack's approach
is highly tuned for stack protection and may not scale against heap
vulnerabilities in its current state. Finally, RA-malloc~\cite{Jang2019:RUMA} permutes 
heap allocations by randomizing the byte location of the starting address of 
pointers. As pointers are of~$8$-byte size, RA-malloc offers~$\left(\frac{1}{8}\right)$ success 
probability for the attacker compared to~$\left(\frac{1}{16!}\right)$ for~\pname{}.

\fakesec{Secure Allocators.} 
Secure memory allocators provide probabilistic guarantees for spatial memory 
protection, temporal memory 
protection (\eg{} FreeSentry~\cite{Younan2015:FreeSentry} and 
Oscar~\cite{Dang2017:Oscar}), or both (\eg{} DieHard~\cite{Berger2006:DieHard}, DieHarder~\cite{Emery2010:dieharder}, 
FreeGuard~\cite{Silvestro2017:FreeGuard}, and GUARDER~\cite{sam2018:Guarder}). 
Although those solutions typically come with minimal performance overheads, they 
lack protection against intra-object memory violations. Additionally, they only 
randomize the starting location of the newly allocated object; making them 
vulnerable to hardware memory corruption attacks unlike~\pname{} that permutes the 
object-data itself.

\section{Conclusion}\label{sec:conclusion}

In this paper, we presented~\pname{}, a software defense that significantly 
improves the resilience of applications to software and hardware memory 
violations.~\pname{}'s novel insight of permuting data based on its memory location 
permits per-instance dynamic permutation. Our proposed~\b2p{} transformation 
allows~\pname{} to guarantee sub-object protection against data corruption 
attacks. We built an initial prototype to demonstrate~\pname{} using an LLVM 
compiler pass with an extension to to the \texttt{compiler-rt} runtime. We successfully 
compiled and ran a variety of applications with~\pname{} to show 
its deployability.~\pname{} + \b2p{} provides comprehensive  
security guarantees and has modest overheads for I/O bound workloads (\eg{}
$1.4$x for Nginx). Moreover,~\pname{} efficiently scales with 
multi-threaded applications. Our security evaluation shows that~\pname{} 
provides strong probabilistic protection (where an attacker chance of success is as 
low as~$\frac{1}{16!} \approx 10^{-14}$) against a wide range of memory corruption 
attacks that have not been previously covered by a single mitigation. 

Our experience developing~\pname{} shows that further optimizations would allow
the security benefits of data permutation to extend to other parts of a system:
(1) a~\pname{} instrumented OS can significantly harden the root of trust (2) a
specialized \pname{} allocator would allow us to provide fine-grained memory safety
to~$32$-bit systems without high runtime overheads (3) while we currently
support C programs, with more engineering effort we can support C++ programs as
well and (4) minimal hardware extensions can significantly reduce the overhead making
\pname{} applicable for a wide variety of workloads. We believe that cohesive
solutions like~\pname{} are mandatory to stand against the daily influx of
attacks.

\bibliographystyle{IEEEtranS}
\bibliography{main}

% Generated by IEEEtranS.bst, version: 1.12 (2007/01/11)
\begin{thebibliography}{10}
\providecommand{\url}[1]{#1}
\csname url@samestyle\endcsname
\providecommand{\newblock}{\relax}
\providecommand{\bibinfo}[2]{#2}
\providecommand{\BIBentrySTDinterwordspacing}{\spaceskip=0pt\relax}
\providecommand{\BIBentryALTinterwordstretchfactor}{4}
\providecommand{\BIBentryALTinterwordspacing}{\spaceskip=\fontdimen2\font plus
\BIBentryALTinterwordstretchfactor\fontdimen3\font minus
  \fontdimen4\font\relax}
\providecommand{\BIBforeignlanguage}[2]{{%
\expandafter\ifx\csname l@#1\endcsname\relax
\typeout{** WARNING: IEEEtranS.bst: No hyphenation pattern has been}%
\typeout{** loaded for the language `#1'. Using the pattern for}%
\typeout{** the default language instead.}%
\else
\language=\csname l@#1\endcsname
\fi
#2}}
\providecommand{\BIBdecl}{\relax}
\BIBdecl

\bibitem{Aga2019:Smokestack}
M.~T. Aga and T.~Austin, ``Smokestack: Thwarting {DOP} attacks with runtime
  stack layout randomization,'' in \emph{Proceedings of the 2019 IEEE/ACM
  International Symposium on Code Generation and Optimization}, ser. CGO 2019,
  Washington, DC, USA, 2019, pp. 26--36.

\bibitem{Ahmad2015:CRONO}
M.~{Ahmad}, F.~{Hijaz}, Q.~{Shi}, and O.~{Khan}, ``{CRONO}: A benchmark suite
  for multithreaded graph algorithms executing on futuristic multicores,'' in
  \emph{2015 IEEE International Symposium on Workload Characterization},
  Atlanta, GA, USA, Oct 2015, pp. 44--55.

\bibitem{ARM2019:MTE}
\BIBentryALTinterwordspacing
{ARM}, ``Memory tagging extension: Enhancing memory safety through
  architecture,'' 2019. [Online]. Available:
  \url{https://community.arm.com/developer/ip-products/processors/b/processors-ip-blog/posts/enhancing-memory-safety}
\BIBentrySTDinterwordspacing

\bibitem{Todd1995:PtrDist}
\BIBentryALTinterwordspacing
T.~Austin, ``Pointer-intensive benchmark suite,'' Sept 1995. [Online].
  Available: \url{http://pages.cs.wisc.edu/~austin/ptr-dist.html}
\BIBentrySTDinterwordspacing

\bibitem{Berger2006:DieHard}
E.~D. Berger and B.~G. Zorn, ``Diehard: Probabilistic memory safety for unsafe
  languages,'' in \emph{Proceedings of the 27th ACM SIGPLAN Conference on
  Programming Language Design and Implementation}, ser. PLDI '06, Ottawa,
  Ontario, Canada, 2006, pp. 158--168.

\bibitem{Bhatkar2008:DSR}
S.~Bhatkar and R.~Sekar, ``Data space randomization,'' in \emph{Proceedings of
  the 5th International Conference on Detection of Intrusions and Malware, and
  Vulnerability Assessment}, ser. DIMVA '08.\hskip 1em plus 0.5em minus
  0.4em\relax Paris, France: Springer-Verlag, 2008, pp. 1--22.

\bibitem{Bletsch2011:JOP}
T.~Bletsch, X.~Jiang, V.~W. Freeh, and Z.~Liang, ``Jump-oriented programming: A
  new class of code-reuse attack,'' in \emph{Proceedings of the 6th ACM
  Symposium on Information, Computer and Communications Security}, ser. ASIACCS
  '11, Hong Kong, China, 2011, pp. 30--40.

\bibitem{Bucek2018:SPEC}
J.~Bucek, K.-D. Lange, and J.~v.~Kistowski, ``{SPEC CPU2017}: Next-generation
  compute benchmark,'' in \emph{Proceedings of the 2018 ACM/SPEC International
  Conference on Performance Engineering}, ser. ICPE '18, April 2018, pp.
  41--42.

\bibitem{Burow2018:CUP}
\BIBentryALTinterwordspacing
N.~Burow, D.~McKee, S.~A. Carr, and M.~Payer, ``{CUP}: Comprehensive user-space
  protection for {C/C++},'' in \emph{Proceedings of the 2018 on Asia Conference
  on Computer and Communications Security}, ser. ASIACCS ’18.\hskip 1em plus
  0.5em minus 0.4em\relax New York, NY, USA: Association for Computing
  Machinery, 2018, p. 381–392. [Online]. Available:
  \url{https://doi.org/10.1145/3196494.3196540}
\BIBentrySTDinterwordspacing

\bibitem{DBLP:conf/osdi/CastroCH06}
\BIBentryALTinterwordspacing
M.~Castro, M.~Costa, and T.~Harris, ``Securing software by enforcing data-flow
  integrity,'' in \emph{7th Symposium on Operating Systems Design and
  Implementation {(OSDI} '06), November 6-8, Seattle, WA, {USA}}, B.~N. Bershad
  and J.~C. Mogul, Eds.\hskip 1em plus 0.5em minus 0.4em\relax {USENIX}
  Association, 2006, pp. 147--160. [Online]. Available:
  \url{http://www.usenix.org/events/osdi06/tech/castro.html}
\BIBentrySTDinterwordspacing

\bibitem{Chen2018:saladsPlus}
P.~Chen, Z.~Hu, J.~Xu, M.~Zhu, and P.~Liu, ``Feedback control can make data
  structure layout randomization more cost-effective under zero-day attacks,''
  \emph{Cybersecurity}, vol.~1, no.~1, p.~3, Jun 2018.

\bibitem{Chen2015:salads}
P.~Chen, J.~Xu, Z.~Lin, D.~Xu, B.~Mao, and P.~Liu, ``A practical approach for
  adaptive data structure layout randomization,'' in \emph{Computer Security --
  ESORICS 2015}, G.~Pernul, P.~Y~A~Ryan, and E.~Weippl, Eds.\hskip 1em plus
  0.5em minus 0.4em\relax Cham: Springer International Publishing, 2015, pp.
  69--89.

\bibitem{Chen2005:Non-control}
S.~Chen, J.~Xu, E.~C. Sezer, P.~Gauriar, and R.~K. Iyer, ``Non-control-data
  attacks are realistic threats,'' in \emph{Proceedings of the 14th Conference
  on USENIX Security Symposium - Volume 14}, ser. SSYM'05.\hskip 1em plus 0.5em
  minus 0.4em\relax Baltimore, MD, USA: USENIX Association, 2005, pp. 12--12.

\bibitem{Cheng2019:SoKDOP}
L.~{Cheng}, H.~{Liljestrand}, M.~S. {Ahmed}, T.~{Nyman}, T.~{Jaeger},
  N.~{Asokan}, and D.~{Yao}, ``Exploitation techniques and defenses for
  data-oriented attacks,'' in \emph{Proceedings of the 2019 IEEE Cybersecurity
  Development (SecDev)}, Tysons Corner, VA, USA, September 2019, pp. 114--128.

\bibitem{Costa2008:MicrosoftDR}
\BIBentryALTinterwordspacing
M.~Costa, J.-P. Martin, and M.~Castro, ``Data randomization,'' Microsoft, Tech.
  Rep. MSR-TR-2008-120, September 2008. [Online]. Available:
  \url{https://www.microsoft.com/en-us/research/publication/data-randomization/}
\BIBentrySTDinterwordspacing

\bibitem{Crispin2003:Pointguard}
C.~Cowan, S.~Beattie, J.~Johansen, and P.~Wagle, ``Pointguard: Protecting
  pointers from buffer overflow vulnerabilities,'' in \emph{Proceedings of the
  12th Conference on USENIX Security Symposium - Volume 12}, ser.
  SSYM’03.\hskip 1em plus 0.5em minus 0.4em\relax Washington, DC, USA: USENIX
  Association, 2003, p.~7.

\bibitem{Emery2013:STABILIZER}
\BIBentryALTinterwordspacing
C.~Curtsinger and E.~D. Berger, ``Stabilizer: Statistically sound performance
  evaluation,'' in \emph{Proceedings of the Eighteenth International Conference
  on Architectural Support for Programming Languages and Operating Systems},
  ser. ASPLOS ’13.\hskip 1em plus 0.5em minus 0.4em\relax Houston, Texas,
  USA: Association for Computing Machinery, 2013, p. 219–228. [Online].
  Available: \url{https://doi.org/10.1145/2451116.2451141}
\BIBentrySTDinterwordspacing

\bibitem{Dang2017:Oscar}
\BIBentryALTinterwordspacing
T.~H. Dang, P.~Maniatis, and D.~Wagner, ``Oscar: A practical
  page-permissions-based scheme for thwarting dangling pointers,'' in
  \emph{26th {USENIX} Security Symposium ({USENIX} Security 17)}.\hskip 1em
  plus 0.5em minus 0.4em\relax Vancouver, BC: {USENIX} Association, Aug. 2017,
  pp. 815--832. [Online]. Available:
  \url{https://www.usenix.org/conference/usenixsecurity17/technical-sessions/presentation/dang}
\BIBentrySTDinterwordspacing

\bibitem{isomeron2015}
L.~Davi, C.~Liebchen, A.-R. Sadeghi, K.~Z.~Snow, and F.~Monrose, ``Isomeron:
  Code randomization resilient to ({Just-In-Time}) return-oriented
  programming,'' in \emph{Proceedings of the 2015 Network and Distributed
  System Security Symposium}, ser. NDSS~'15, San Diego, CA, USA, February 2015.

\bibitem{PractRand}
\BIBentryALTinterwordspacing
C.~Doty-Humphrey, ``Practically random.'' [Online]. Available:
  \url{http://pracrand.sourceforge.net/}
\BIBentrySTDinterwordspacing

\bibitem{Duck2016:LowFatSW}
\BIBentryALTinterwordspacing
G.~J. Duck and R.~H.~C. Yap, ``Heap bounds protection with low fat pointers,''
  in \emph{Proceedings of the 25th International Conference on Compiler
  Construction}, ser. CC~'16.\hskip 1em plus 0.5em minus 0.4em\relax Barcelona,
  Spain: ACM, 2016, pp. 132--142. [Online]. Available:
  \url{http://doi.acm.org/10.1145/2892208.2892212}
\BIBentrySTDinterwordspacing

\bibitem{Duck2018:effectiveSAN}
G.~J. Duck and R.~H.~C. Yap, ``{EffectiveSan}: type and memory error detection
  using dynamically typed {C/C++},'' in \emph{PLDI~'18: Proceedings of the 39th
  ACM SIGPLAN Conference on Programming Language Design and Implementation},
  2018.

\bibitem{duck2018extended}
G.~J. Duck and R.~H.~C. Yap, ``An extended low fat allocator api and
  applications,'' 2018.

\bibitem{duck2018stack}
\BIBentryALTinterwordspacing
G.~J. Duck, R.~H.~C. Yap, and L.~Cavallaro, ``Stack bounds protection with low
  fat pointers,'' in \emph{24th Annual Network and Distributed System Security
  Symposium, {NDSS} 2017, San Diego, California, USA, February 26 - March 1,
  2017}.\hskip 1em plus 0.5em minus 0.4em\relax The Internet Society, 2017.
  [Online]. Available:
  \url{https://www.ndss-symposium.org/ndss2017/ndss-2017-programme/stack-object-protection-low-fat-pointers/}
\BIBentrySTDinterwordspacing

\bibitem{10.1145/364520.364540}
\BIBentryALTinterwordspacing
R.~Durstenfeld, ``Algorithm 235: Random permutation,'' \emph{Commun. ACM},
  vol.~7, no.~7, p. 420, Jul. 1964. [Online]. Available:
  \url{https://doi.org/10.1145/364520.364540}
\BIBentrySTDinterwordspacing

\bibitem{CWE:FormatString}
\BIBentryALTinterwordspacing
C.~W. Enumeration, ``{CWE-134}: Use of externally-controlled format string.''
  [Online]. Available: \url{https://cwe.mitre.org/data/definitions/134.html}
\BIBentrySTDinterwordspacing

\bibitem{FisherYates}
\BIBentryALTinterwordspacing
S.~R.~A. Fisher and F.~Yates, ``Statistical tables for biological, agricultural
  and medical research,'' 1963. [Online]. Available:
  \url{http://hdl.handle.net/2440/10701}
\BIBentrySTDinterwordspacing

\bibitem{siege}
J.~Fulmer, ``Siege,'' \url{https://github.com/JoeDog/siege/}, 2012.

\bibitem{octane2}
Google, ``Octane 2,'' \url{https://github.com/chromium/octane}, 2012.

\bibitem{projectzero2015:confusion}
\BIBentryALTinterwordspacing
GoogleProjectZero, ``One perfect bug: Exploiting type confusion in flash,''
  2015. [Online]. Available:
  \url{https://googleprojectzero.blogspot.com/2015/07/one-perfect-bug-exploiting-type_20.html}
\BIBentrySTDinterwordspacing

\bibitem{projectzero2019:0day}
\BIBentryALTinterwordspacing
GoogleProjectZero, ``0day in the wild,'' 2019. [Online]. Available:
  \url{https://googleprojectzero.blogspot.com/p/0day.html}
\BIBentrySTDinterwordspacing

\bibitem{DBLP:conf/fse/GranboulanP07}
\BIBentryALTinterwordspacing
L.~Granboulan and T.~Pornin, ``Perfect block ciphers with small blocks,'' in
  \emph{Fast Software Encryption, 14th International Workshop, {FSE} 2007,
  Luxembourg, Luxembourg, March 26-28, 2007, Revised Selected Papers}, ser.
  Lecture Notes in Computer Science, A.~Biryukov, Ed., vol. 4593.\hskip 1em
  plus 0.5em minus 0.4em\relax Springer, 2007, pp. 452--465. [Online].
  Available: \url{https://doi.org/10.1007/978-3-540-74619-5\_28}
\BIBentrySTDinterwordspacing

\bibitem{Gruss2015:cache}
\BIBentryALTinterwordspacing
D.~Gruss, R.~Spreitzer, and S.~Mangard, ``Cache template attacks: Automating
  attacks on inclusive last-level caches,'' in \emph{24th {USENIX} Security
  Symposium ({USENIX} Security 15)}.\hskip 1em plus 0.5em minus 0.4em\relax
  Washington, D.C.: {USENIX} Association, Aug. 2015, pp. 897--912. [Online].
  Available:
  \url{https://www.usenix.org/conference/usenixsecurity15/technical-sessions/presentation/gruss}
\BIBentrySTDinterwordspacing

\bibitem{Marcus2017:side}
\BIBentryALTinterwordspacing
M.~H{\"a}hnel, W.~Cui, and M.~Peinado, ``High-resolution side channels for
  untrusted operating systems,'' in \emph{2017 {USENIX} Annual Technical
  Conference ({USENIX} {ATC} 17)}.\hskip 1em plus 0.5em minus 0.4em\relax Santa
  Clara, CA: {USENIX} Association, Jul. 2017, pp. 299--312. [Online].
  Available:
  \url{https://www.usenix.org/conference/atc17/technical-sessions/presentation/hahnel}
\BIBentrySTDinterwordspacing

\bibitem{Halderman2009:ColdBoot}
J.~A. Halderman, S.~D. Schoen, N.~Heninger, W.~Clarkson, W.~Paul, J.~A.
  Calandrino, A.~J. Feldman, J.~Appelbaum, and E.~W. Felten, ``Lest we
  remember: cold-boot attacks on encryption keys,'' \emph{Communications of the
  ACM}, vol.~52, no.~5, pp. 91--98, 2009.

\bibitem{Hodor:2019}
\BIBentryALTinterwordspacing
M.~Hedayati, S.~Gravani, E.~Johnson, J.~Criswell, M.~L. Scott, K.~Shen, and
  M.~Marty, ``Hodor: Intra-process isolation for high-throughput data plane
  libraries,'' in \emph{2019 {USENIX} Annual Technical Conference ({USENIX}
  {ATC} 19)}.\hskip 1em plus 0.5em minus 0.4em\relax Renton, WA: {USENIX}
  Association, Jul. 2019, pp. 489--504. [Online]. Available:
  \url{https://www.usenix.org/conference/atc19/presentation/hedayati-hodor}
\BIBentrySTDinterwordspacing

\bibitem{DBLP:conf/crypto/HoangMR12}
\BIBentryALTinterwordspacing
V.~T. Hoang, B.~Morris, and P.~Rogaway, ``An enciphering scheme based on a card
  shuffle,'' in \emph{Advances in Cryptology - {CRYPTO} 2012 - 32nd Annual
  Cryptology Conference, Santa Barbara, CA, USA, August 19-23, 2012.
  Proceedings}, ser. Lecture Notes in Computer Science, R.~Safavi{-}Naini and
  R.~Canetti, Eds., vol. 7417.\hskip 1em plus 0.5em minus 0.4em\relax Springer,
  2012, pp. 1--13. [Online]. Available:
  \url{https://doi.org/10.1007/978-3-642-32009-5\_1}
\BIBentrySTDinterwordspacing

\bibitem{httparchive}
{HTTP}, ``The 2019 {Web Almanac}: Http archive's annual state of the web
  report,'' \url{https://almanac.httparchive.org/en/2019/}, 2019.

\bibitem{Hu2016:DOP}
H.~Hu, S.~Shinde, S.~Adrian, Z.~L. Chua, P.~Saxena, and Z.~Liang,
  ``Data-oriented programming: On the expressiveness of non-control data
  attacks,'' in \emph{2016 IEEE Symposium on Security and Privacy (SP)}, San
  Jose, CA, USA, May 2016, pp. 969--986.

\bibitem{Hundt2006:DLO}
\BIBentryALTinterwordspacing
R.~Hundt, S.~Mannarswamy, and D.~Chakrabarti, ``Practical structure layout
  optimization and advice,'' in \emph{Proceedings of the International
  Symposium on Code Generation and Optimization}, ser. CGO ’06.\hskip 1em
  plus 0.5em minus 0.4em\relax USA: IEEE Computer Society, 2006, p. 233244.
  [Online]. Available: \url{https://doi.org/10.1109/CGO.2006.29}
\BIBentrySTDinterwordspacing

\bibitem{IntelCeT}
\BIBentryALTinterwordspacing
Intel, ``Intel control-flow enforcement technology preview,'' 2017. [Online].
  Available:
  \url{https://software.intel.com/sites/default/files/managed/4d/2a/control-flow-enforcement-technology-preview.pdf}
\BIBentrySTDinterwordspacing

\bibitem{ISO24731}
{ISO}, ``{Information technology -- Programming languages, their environments
  and system software interfaces -- Extensions to the C library -- Part 2:
  Dynamic Allocation Functions},'' International Organization for
  Standardization, Geneva, CH, Standard, Dec. 2010.

\bibitem{Jang2019:RUMA}
D.~Jang, J.~Kim, H.~Lee, M.~Park, Y.~Jung, M.~Kim, and B.~B. Kang, ``On the
  analysis of byte-granularity heap randomization,'' \emph{IEEE Transactions on
  Dependable and Secure Computing}, pp. 1--1, 2019.

\bibitem{Kim2019:Polar}
J.~Kim, D.~Jang, Y.~Jeong, and B.~B. Kang, ``{POLaR}: Per-allocation object
  layout randomization,'' in \emph{Proceedings of the 2019 IEEE/IFIP
  International Conference on Dependable Systems and Networks}, ser. DSN 2019,
  Portland, Oregon, USA, June 2019.

\bibitem{Kim2014:RowHammer}
Y.~Kim, R.~Daly, J.~Kim, C.~Fallin, J.~H. Lee, D.~Lee, C.~Wilkerson, K.~Lai,
  and O.~Mutlu, ``Flipping bits in memory without accessing them: An
  experimental study of dram disturbance errors,'' in \emph{Proceeding of the
  41st Annual International Symposium on Computer Architecuture}, ser. ISCA
  ’14.\hskip 1em plus 0.5em minus 0.4em\relax IEEE Press, 2014, p. 361–372.

\bibitem{10.5555/270146}
D.~E. Knuth, \emph{The Art of Computer Programming, Volume 2 (3rd Ed.):
  Seminumerical Algorithms}.\hskip 1em plus 0.5em minus 0.4em\relax USA:
  Addison-Wesley Longman Publishing Co., Inc., 1997.

\bibitem{Kocher2018:spectre}
P.~Kocher, D.~Genkin, D.~Gruss, W.~Haas, M.~Hamburg, M.~Lipp, S.~Mangard,
  T.~Prescher, M.~Schwarz, and Y.~Yarom, ``Spectre attacks: Exploiting
  speculative execution,'' in \emph{40th IEEE Symposium on Security and Privacy
  (S\&P'19)}, May 2019.

\bibitem{BigCrush}
P.~L'Ecuyer and R.~Simard, ``{TestU01}: A {C} library for empirical testing of
  random number generators,'' \emph{ACM Transactions on Mathematical Software
  (TOMS)}, vol.~33, no.~4, pp. 1--40, 2007.

\bibitem{Lee2001:permute}
R.~B. {Lee}, {Zhijie Shi}, and {Xiiao Yang}, ``Efficient permutation
  instructions for fast software cryptography,'' \emph{IEEE Micro}, vol.~21,
  no.~6, pp. 56--69, Nov 2001.

\bibitem{rand}
\BIBentryALTinterwordspacing
G.~Library, ``{ISO C} random number functions.'' [Online]. Available:
  \url{https://www.gnu.org/software/libc/manual/html_node/ISO-Random.html#ISO-Random}
\BIBentrySTDinterwordspacing

\bibitem{Hans2019:PAC}
\BIBentryALTinterwordspacing
H.~Liljestrand, T.~Nyman, K.~Wang, C.~C. Perez, J.-E. Ekberg, and N.~Asokan,
  ``{PAC} it up: Towards pointer integrity using {ARM} pointer
  authentication,'' in \emph{28th {USENIX} Security Symposium ({USENIX}
  Security 19)}.\hskip 1em plus 0.5em minus 0.4em\relax Santa Clara, CA:
  {USENIX} Association, Aug. 2019, pp. 177--194. [Online]. Available:
  \url{https://www.usenix.org/conference/usenixsecurity19/presentation/liljestrand}
\BIBentrySTDinterwordspacing

\bibitem{Lin2009:polymorphing}
Z.~Lin, R.~D. Riley, and D.~Xu, ``Polymorphing software by randomizing data
  structure layout,'' in \emph{Detection of Intrusions and Malware, and
  Vulnerability Assessment}, U.~Flegel and D.~Bruschi, Eds.\hskip 1em plus
  0.5em minus 0.4em\relax Springer Berlin Heidelberg, 2009, pp. 107--126.

\bibitem{xorshift}
\BIBentryALTinterwordspacing
G.~Marsaglia, ``Xorshift {RNGs},'' \emph{Journal of Statistical Software,
  Articles}, vol.~8, no.~14, pp. 1--6, 2003. [Online]. Available:
  \url{https://www.jstatsoft.org/v008/i14}
\BIBentrySTDinterwordspacing

\bibitem{Nagarakatte2009:SoftBound}
\BIBentryALTinterwordspacing
S.~Nagarakatte, J.~Zhao, M.~M. Martin, and S.~Zdancewic, ``{SoftBound}: Highly
  compatible and complete spatial memory safety for {C},'' in \emph{Proceedings
  of the 30th ACM SIGPLAN Conference on Programming Language Design and
  Implementation}, ser. PLDI '09.\hskip 1em plus 0.5em minus 0.4em\relax
  Dublin, Ireland: ACM, 2009, pp. 245--258. [Online]. Available:
  \url{http://doi.acm.org/10.1145/1542476.1542504}
\BIBentrySTDinterwordspacing

\bibitem{Nagarakatte2010:CETS}
\BIBentryALTinterwordspacing
S.~Nagarakatte, J.~Zhao, M.~M. Martin, and S.~Zdancewic, ``{CETS}: Compiler
  enforced temporal safety for {C},'' in \emph{Proceedings of the 2010
  International Symposium on Memory Management}, ser. ISMM '10.\hskip 1em plus
  0.5em minus 0.4em\relax Toronto, Ontario, Canada: ACM, 2010, pp. 31--40.
  [Online]. Available: \url{http://doi.acm.org/10.1145/1806651.1806657}
\BIBentrySTDinterwordspacing

\bibitem{Emery2010:dieharder}
\BIBentryALTinterwordspacing
G.~Novark and E.~D. Berger, ``Dieharder: Securing the heap,'' in
  \emph{Proceedings of the 17th ACM Conference on Computer and Communications
  Security}, ser. CCS ’10.\hskip 1em plus 0.5em minus 0.4em\relax Chicago,
  Illinois, USA: Association for Computing Machinery, 2010, p. 573–584.
  [Online]. Available: \url{https://doi.org/10.1145/1866307.1866371}
\BIBentrySTDinterwordspacing

\bibitem{mcsema}
\BIBentryALTinterwordspacing
T.~of~Bits, ``Mcsema,'' 2014. [Online]. Available:
  \url{https://github.com/lifting-bits/mcsema}
\BIBentrySTDinterwordspacing

\bibitem{oleksenko2018intel}
O.~Oleksenko, D.~Kuvaiskii, P.~Bhatotia, P.~Felber, and C.~Fetzer, ``Intel mpx
  explained: A cross-layer analysis of the intel mpx system stack,''
  \emph{Proceedings of the ACM on Measurement and Analysis of Computing
  Systems}, vol.~2, no.~2, p.~28, 2018.

\bibitem{Park2019:libmpk}
\BIBentryALTinterwordspacing
S.~Park, S.~Lee, W.~Xu, H.~Moon, and T.~Kim, ``libmpk: Software abstraction for
  intel memory protection keys (intel {MPK}),'' in \emph{2019 {USENIX} Annual
  Technical Conference ({USENIX} {ATC} 19)}.\hskip 1em plus 0.5em minus
  0.4em\relax Renton, WA: {USENIX} Association, Jul. 2019, pp. 241--254.
  [Online]. Available:
  \url{https://www.usenix.org/conference/atc19/presentation/park-soyeon}
\BIBentrySTDinterwordspacing

\bibitem{Pewny2019:STEROIDS}
J.~Pewny, P.~Koppe, and T.~Holz, ``{STEROIDS} for {DOPed} applications: A
  compiler for automated data-oriented programming,'' in \emph{2019 IEEE
  European Symposium on Security and Privacy (EuroS\&P)}, Stockholm, Sweden,
  June 2019.

\bibitem{Emery2019:MESH}
\BIBentryALTinterwordspacing
B.~Powers, D.~Tench, E.~D. Berger, and A.~McGregor, ``Mesh: Compacting memory
  management for c/c++ applications,'' in \emph{Proceedings of the 40th ACM
  SIGPLAN Conference on Programming Language Design and Implementation}, ser.
  PLDI 2019.\hskip 1em plus 0.5em minus 0.4em\relax Phoenix, AZ, USA:
  Association for Computing Machinery, 2019, p. 333–346. [Online]. Available:
  \url{https://doi.org/10.1145/3314221.3314582}
\BIBentrySTDinterwordspacing

\bibitem{Qualcomm2017}
\BIBentryALTinterwordspacing
I.~Qualcomm~Technologies, ``Pointer authentication on {ARMv8.3},'' 2017.
  [Online]. Available:
  \url{https://www.qualcomm.com/media/documents/files/whitepaper-pointer-authentication-on-armv8-3.pdf}
\BIBentrySTDinterwordspacing

\bibitem{DBLP:conf/crypto/RistenpartY13}
\BIBentryALTinterwordspacing
T.~Ristenpart and S.~Yilek, ``The mix-and-cut shuffle: Small-domain encryption
  secure against {N} queries,'' in \emph{Advances in Cryptology - {CRYPTO} 2013
  - 33rd Annual Cryptology Conference, Santa Barbara, CA, USA, August 18-22,
  2013. Proceedings, Part {I}}, ser. Lecture Notes in Computer Science,
  R.~Canetti and J.~A. Garay, Eds., vol. 8042.\hskip 1em plus 0.5em minus
  0.4em\relax Springer, 2013, pp. 392--409. [Online]. Available:
  \url{https://doi.org/10.1007/978-3-642-40041-4\_22}
\BIBentrySTDinterwordspacing

\bibitem{Rogaway10asynopsis}
P.~Rogaway, ``A synopsis of format-preserving encryption,'' 2010.

\bibitem{Anne1995:Olden}
A.~Rogers, M.~C. Carlisle, J.~H. Reppy, and L.~J. Hendren, ``Supporting dynamic
  data structures on distributed-memory machines,'' \emph{ACM Trans. Program.
  Lang. Syst.}, vol.~17, no.~2, p. 233–263, Mar. 1995.

\bibitem{Roy2016:StructSlim}
\BIBentryALTinterwordspacing
P.~Roy and X.~Liu, ``Structslim: A lightweight profiler to guide structure
  splitting,'' in \emph{Proceedings of the 2016 International Symposium on Code
  Generation and Optimization}, ser. CGO ’16.\hskip 1em plus 0.5em minus
  0.4em\relax New York, NY, USA: Association for Computing Machinery, 2016, p.
  36–46. [Online]. Available: \url{https://doi.org/10.1145/2854038.2854053}
\BIBentrySTDinterwordspacing

\bibitem{Saltzer1975}
J.~H. {Saltzer} and M.~D. {Schroeder}, ``The protection of information in
  computer systems,'' \emph{Proceedings of the IEEE}, vol.~63, no.~9, pp.
  1278--1308, Sep. 1975.

\bibitem{Sasaki2019:Califorms}
\BIBentryALTinterwordspacing
H.~Sasaki, M.~A. Arroyo, M.~T.~I. Ziad, K.~Bhat, K.~Sinha, and
  S.~Sethumadhavan, ``Practical byte-granular memory blacklisting using
  {C}aliforms,'' in \emph{Proceedings of the 52nd Annual IEEE/ACM International
  Symposium on Microarchitecture}, ser. MICRO ’52.\hskip 1em plus 0.5em minus
  0.4em\relax Columbus, OH, USA: Association for Computing Machinery, 2019, p.
  558–571. [Online]. Available: \url{https://doi.org/10.1145/3352460.3358299}
\BIBentrySTDinterwordspacing

\bibitem{Schuster2015:COOP}
F.~Schuster, T.~Tendyck, C.~Liebchen, L.~Davi, A.-R. Sadeghi, and T.~Holz,
  ``Counterfeit object-oriented programming: On the difficulty of preventing
  code reuse attacks in {C++} applications,'' in \emph{Proceedings of the 2015
  IEEE Symposium on Security and Privacy}, ser. SP '15, Oakland, CA, USA, 2015,
  pp. 745--762.

\bibitem{Serebryany2012:ASAN}
K.~Serebryany, D.~Bruening, A.~Potapenko, and D.~Vyukov, ``{AddressSanitizer}:
  a fast address sanity checker,'' in \emph{USENIX ATC~'12: Proceedings of the
  2012 USENIX Annual Technical Conference}, 2012.

\bibitem{Shi2003:two-cycle}
Z.~{Shi}, X.~{Yang}, and R.~B. {Lee}, ``Arbitrary bit permutations in one or
  two cycles,'' in \emph{Proceedings IEEE International Conference on
  Application-Specific Systems, Architectures, and Processors. ASAP 2003}, The
  Hague, Netherlands, June 2003, pp. 237--247.

\bibitem{Silvestro2017:FreeGuard}
\BIBentryALTinterwordspacing
S.~Silvestro, H.~Liu, C.~Crosser, Z.~Lin, and T.~Liu, ``{FreeGuard}: A faster
  secure heap allocator,'' in \emph{Proceedings of the 2017 ACM SIGSAC
  Conference on Computer and Communications Security}, ser. CCS '17.\hskip 1em
  plus 0.5em minus 0.4em\relax Dallas, Texas, USA: ACM, 2017, pp. 2389--2403.
  [Online]. Available: \url{http://doi.acm.org/10.1145/3133956.3133957}
\BIBentrySTDinterwordspacing

\bibitem{sam2018:Guarder}
S.~Silvestro, H.~Liu, T.~Liu, Z.~Lin, and T.~Liu, ``Guarder: A tunable secure
  allocator,'' in \emph{27th {USENIX} Security Symposium ({USENIX} Security
  18)}.\hskip 1em plus 0.5em minus 0.4em\relax Baltimore, MD, USA: {USENIX}
  Association, 2018, pp. 117--133.

\bibitem{snow2013:JIT-CRA}
K.~Z. Snow, F.~Monrose, L.~Davi, A.~Dmitrienko, C.~Liebchen, and A.-R. Sadeghi,
  ``Just-in-time code reuse: On the effectiveness of fine-grained address space
  layout randomization,'' in \emph{2013 IEEE Symposium on Security and
  Privacy}, Berkeley, CA, USA, May 2013, pp. 574--588.

\bibitem{Sotirov2007:heapfengshui}
A.~Sotirov, ``Heap feng shui in javascript,'' \emph{BlackHat}, 2007.

\bibitem{Stanley2013}
D.~M. Stanley, D.~Xu, and E.~H. Spafford, ``Improved kernel security through
  memory layout randomization,'' in \emph{2013 IEEE 32nd International
  Performance Computing and Communications Conference (IPCCC)}, San Diego, CA,
  USA, Dec 2013, pp. 1--10.

\bibitem{Raoul2009:Leak}
\BIBentryALTinterwordspacing
R.~Strackx, Y.~Younan, P.~Philippaerts, F.~Piessens, S.~Lachmund, and
  T.~Walter, ``Breaking the memory secrecy assumption,'' in \emph{Proceedings
  of the Second European Workshop on System Security}, ser. EUROSEC
  ’09.\hskip 1em plus 0.5em minus 0.4em\relax Nuremburg, Germany: Association
  for Computing Machinery, 2009, p. 1–8. [Online]. Available:
  \url{https://doi.org/10.1145/1519144.1519145}
\BIBentrySTDinterwordspacing

\bibitem{nginx}
I.~Sysoev, ``Nginx,'' \url{http://nginx.org/en/}, 2004.

\bibitem{AESRand}
\BIBentryALTinterwordspacing
P.~D. Tiglao, ``Aesrand,'' 2018. [Online]. Available:
  \url{https://github.com/dragontamer/AESRand}
\BIBentrySTDinterwordspacing

\bibitem{LinuxMPX}
\BIBentryALTinterwordspacing
L.~Torvalds, ``mpx-for-linus removal,'' Available from Linux Kernel Source
  Tree, 2020. [Online]. Available:
  \url{https://git.kernel.org/pub/scm/linux/kernel/git/torvalds/linux.git/commit/?id=ccaaaf6fe5a5e1fffca5cca0f3fc4ec84d7ae752}
\BIBentrySTDinterwordspacing

\bibitem{duktape}
S.~Vaarala, ``Duktape,'' \url{https://github.com/svaarala/duktape}, 2013.

\bibitem{ERIM:2019}
\BIBentryALTinterwordspacing
A.~Vahldiek-Oberwagner, E.~Elnikety, N.~O. Duarte, M.~Sammler, P.~Druschel, and
  D.~Garg, ``{ERIM}: Secure, efficient in-process isolation with protection
  keys ({MPK}),'' in \emph{28th {USENIX} Security Symposium ({USENIX} Security
  19)}.\hskip 1em plus 0.5em minus 0.4em\relax Santa Clara, CA: {USENIX}
  Association, Aug. 2019, pp. 1221--1238. [Online]. Available:
  \url{https://www.usenix.org/conference/usenixsecurity19/presentation/vahldiek-oberwagner}
\BIBentrySTDinterwordspacing

\bibitem{vanderVeen2017:10Years}
V.~van~der Veen, D.~Andriesse, M.~Stamatogiannakis, X.~Chen, H.~Bos, and
  C.~Giuffrdia, ``The dynamics of innocent flesh on the bone: Code reuse ten
  years later,'' in \emph{Proceedings of the 2017 ACM SIGSAC Conference on
  Computer and Communications Security}, ser. CCS '17, Dallas, Texas, USA,
  2017, pp. 1675--1689.

\bibitem{Wenhao2017}
\BIBentryALTinterwordspacing
W.~Wang, G.~Chen, X.~Pan, Y.~Zhang, X.~Wang, V.~Bindschaedler, H.~Tang, and
  C.~A. Gunter, ``Leaky cauldron on the dark land: Understanding memory
  side-channel hazards in sgx,'' in \emph{Proceedings of the 2017 ACM SIGSAC
  Conference on Computer and Communications Security}, ser. CCS ’17.\hskip
  1em plus 0.5em minus 0.4em\relax Dallas, Texas, USA: Association for
  Computing Machinery, 2017, p. 2421–2434. [Online]. Available:
  \url{https://doi.org/10.1145/3133956.3134038}
\BIBentrySTDinterwordspacing

\bibitem{Wang2020:Shapeshifter}
\BIBentryALTinterwordspacing
Y.~Wang, Q.~Li, Z.~Chen, P.~Zhang, and G.~Zhang, ``Shapeshifter:
  Intelligence-driven data plane randomization resilient to data-oriented
  programming attacks,'' \emph{Computers {\&} Security}, vol.~89, 2020.
  [Online]. Available: \url{https://doi.org/10.1016/j.cose.2019.101679}
\BIBentrySTDinterwordspacing

\bibitem{Watson2015:CHERI}
R.~N.~M. Watson, J.~Woodruff, P.~G. Neumann, S.~W. Moore, J.~Anderson,
  D.~Chisnall, N.~H. Dave, B.~Davis, K.~Gudka, B.~Laurie, S.~J. Murdoch, R.~M.
  Norton, M.~Roe, S.~D. Son, and M.~Vadera, ``{CHERI}: A hybrid
  capability-system architecture for scalable software compartmentalization,''
  in \emph{2015 IEEE Symposium on Security and Privacy}, San Jose, CA, USA, May
  2015, pp. 20--37.

\bibitem{gccMPX}
\BIBentryALTinterwordspacing
G.~Wiki, ``Intel memory protection extensions (intel mpx) support in the gcc
  compiler,'' 2018. [Online]. Available:
  \url{https://gcc.gnu.org/wiki/Intel%20MPX%20support%20in%20the%20GCC%20compiler}
\BIBentrySTDinterwordspacing

\bibitem{Wilander2011:RIPE}
J.~Wilander, N.~Nikiforakis, Y.~Younan, M.~Kamkar, and W.~Joosen, ``{RIPE}:
  Runtime intrusion prevention evaluator,'' in \emph{Proceedings of the 27th
  Annual Computer Security Applications Conference}, ser. ACSAC '11, Orlando,
  Florida, USA, 2011, pp. 41--50.

\bibitem{David2016:shuffler}
D.~Williams-King, G.~Gobieski, K.~Williams-King, J.~P. Blake, X.~Yuan, P.~Colp,
  M.~Zheng, V.~P. Kemerlis, J.~Yang, and W.~Aiello, ``Shuffler: Fast and
  deployable continuous code re-randomization,'' in \emph{Proceedings of the
  12th {USENIX} Symposium on Operating Systems Design and Implementation}, ser.
  OSDI~16.\hskip 1em plus 0.5em minus 0.4em\relax Savannah, GA, USA: USENIX
  Association, 2016, pp. 367--382.

\bibitem{wolfssl}
WolfSSL, ``wolfssl,'' \url{https://github.com/wolfSSL/wolfssl}, 2006.

\bibitem{Hongyan2019:CHERIvoke}
\BIBentryALTinterwordspacing
H.~Xia, J.~Woodruff, S.~Ainsworth, N.~W. Filardo, M.~Roe, A.~Richardson,
  P.~Rugg, P.~G. Neumann, S.~W. Moore, R.~N.~M. Watson, and et~al.,
  ``Cherivoke: Characterising pointer revocation using cheri capabilities for
  temporal memory safety,'' in \emph{Proceedings of the 52nd Annual IEEE/ACM
  International Symposium on Microarchitecture}, ser. MICRO ’52.\hskip 1em
  plus 0.5em minus 0.4em\relax Columbus, OH, USA: Association for Computing
  Machinery, 2019, p. 545–557. [Online]. Available:
  \url{https://doi.org/10.1145/3352460.3358288}
\BIBentrySTDinterwordspacing

\bibitem{Yadavalli2019:mctoll}
\BIBentryALTinterwordspacing
S.~B. Yadavalli and A.~Smith, ``Raising binaries to llvm ir with mctoll (wip
  paper),'' in \emph{Proceedings of the 20th ACM SIGPLAN/SIGBED International
  Conference on Languages, Compilers, and Tools for Embedded Systems}, ser.
  LCTES 2019.\hskip 1em plus 0.5em minus 0.4em\relax New York, NY, USA:
  Association for Computing Machinery, 2019, p. 213–218. [Online]. Available:
  \url{https://doi.org/10.1145/3316482.3326354}
\BIBentrySTDinterwordspacing

\bibitem{Younan2015:FreeSentry}
Y.~Younan, ``{FreeSentry}: Protecting against use-after-free vulnerabilities
  due to dangling pointers,'' in \emph{Proceedings of the 2015 Network and
  Distributed System Security (NDSS) Symposium}, San Diego, CA, USA, February
  2015.

\bibitem{Yu2018:LWP}
C.~Yu, P.~Roy, Y.~Bai, H.~Yang, and X.~Liu, ``{LWPTool}: A lightweight profiler
  to guide data layout optimization,'' \emph{IEEE Transactions on Parallel and
  Distributed Systems}, vol.~29, no.~11, pp. 2489--2502, Nov 2018.

\bibitem{Zhang2019:BOGO}
\BIBentryALTinterwordspacing
T.~Zhang, D.~Lee, and C.~Jung, ``Bogo: Buy spatial memory safety, get temporal
  memory safety (almost) free,'' in \emph{Proceedings of the Twenty-Fourth
  International Conference on Architectural Support for Programming Languages
  and Operating Systems}, ser. ASPLOS ’19.\hskip 1em plus 0.5em minus
  0.4em\relax New York, NY, USA: Association for Computing Machinery, 2019, p.
  631–644. [Online]. Available: \url{https://doi.org/10.1145/3297858.3304017}
\BIBentrySTDinterwordspacing

\end{thebibliography}

\appendices
\section{Additional Evaluation}\label[appendix]{sec:eval-appendix}

This appendix provides more information about our Nginx configuration 
and additional performance results for~\pname{}.    

\subsection{Configuration}

\fakesec{Nginx.} In~\cref{lst:nginx-conf}, we show the configuration file used
for both the baseline and \pname{} enabled Nginx instances used during
evaluation. Here \texttt{benchmark\_html} is a directory that contains the files
of different sizes that are served. The different sized files are generated
using \mintinline{shell}{python -c "print('X'*${FS})" > ${FS}.html} where
\texttt{\$\{FS\}} is the file size. Additionally, we disable non-essential
features as shown in~\cref{lst:nginx-build} to narrow the scope of testing to
just file serving.

\begin{listing}[!h]
  \centering
    \begin{minipage}{0.44\textwidth}
      \begin{minted}
        [
        linenos,
        fontsize=\footnotesize,
        mathescape,
        autogobble,
        stripnl=false,
        numbersep=3pt
        ]
        {nginx}
        worker_processes  1;

        error_log  logs/error.log debug;

        events {
          worker_connections  1024;
        }

        http {
          include       mime.types;
          default_type  application/octet-stream;
          root benchmark_html;
          sendfile        on;
          tcp_nopush on;
          keepalive_timeout  65;

          server {
            listen       8776;
            server_name  localhost;

            location / {
              index index.html;
            }

            error_page   500 502 503 504  /50x.html;
            location = /50x.html {
              root   docs/html;
            }
          }
      }
      \end{minted}
    \end{minipage}
  \caption{The Nginx web server configuration used in our performance evaluation.}\label{lst:nginx-conf}
\end{listing}

\begin{listing}[!h]
  \centering
    \begin{minipage}{0.44\textwidth}
      \begin{minted}
        [
        linenos,
        fontsize=\footnotesize,
        mathescape,
        autogobble,
        stripnl=false,
        numbersep=3pt
        ]
        {shell-session}
        user:~$ auto/configure \
                --without-pcre \
                --without-http_rewrite_module \
                --without-http-cache \
                --without-http_gzip_module \
                --without-http_proxy_module \
                --without-http_limit_conn_module \
                --without-http_limit_req_module \
                --without-http_browser_module \
                --without-http_charset_module \
                --without-http_ssi_module \
                --without-http_userid_module \
                --without-http_access_module \
                --without-http_auth_basic_module \
                --without-http_geo_module \
                --without-http_map_module \
                --without-http_split_clients_module \
                --without-http_referer_module \
                --without-http_rewrite_module \
                --without-http_fastcgi_module \
                --without-http_uwsgi_module \
                --without-http_scgi_module \
                --without-http_memcached_module \
                --without-http_empty_gif_module \
                --without-http_upstream_hash_module \
                --without-http_upstream_ip_hash_module \
                --without-http_upstream_least_conn_module \
                --without-http_upstream_keepalive_module \
                --without-http_upstream_zone_module
      \end{minted}
    \end{minipage}
  \caption{The Nginx build configuration parameters.}\label{lst:nginx-build}
\end{listing}

\subsection{Additional Results}
\begin{figure*}[!h]
  \centering
  \includegraphics[width=0.8\textwidth]{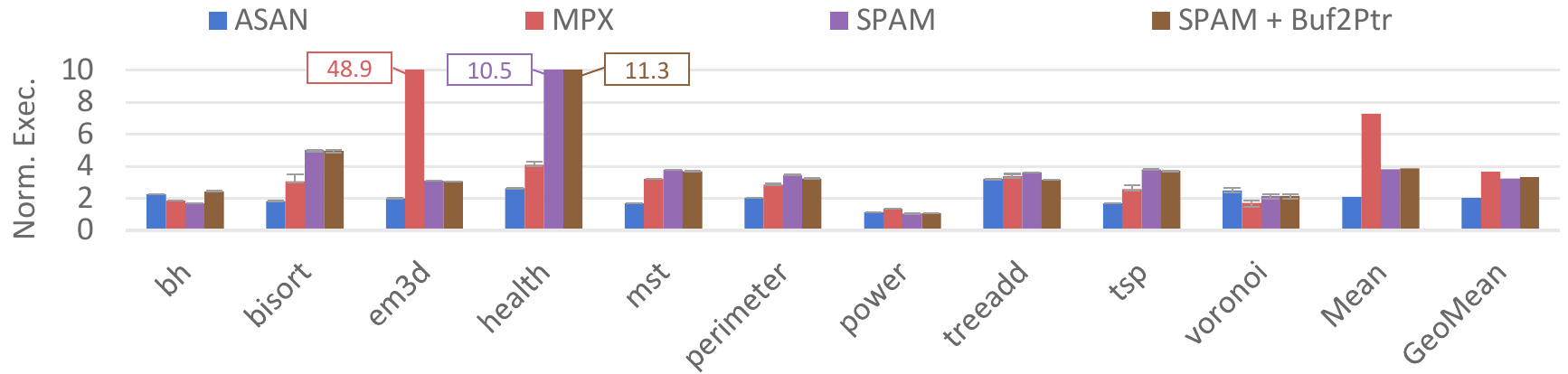}
  \caption{Olden performance for different tools normalized to their
    corresponding baseline for heap memory protection.}\label{fig:olden-perf}
\end{figure*}

\fakesec{Olden.} We further evaluate~\pname{} on a pointer and allocation
intensive benchmark suite, Olden~\cite{Anne1995:Olden}. We use the \texttt{big}
test input as specified by \texttt{llvm-test-suite}.~\cref{fig:olden-perf} shows
the geometric mean of each tool: ASAN ($2.02$x), MPX ($3.65$x), \pname{}
($3.25$x) and \pname{} + \b2p{} ($3.31$x). An interesting outlier for 
MPX is \texttt{em3d} whose singly-linked list data structure stresses the
bounds checking mechanism.

\begin{figure}[!h]
  \centering
  \includegraphics[width=0.45\textwidth]{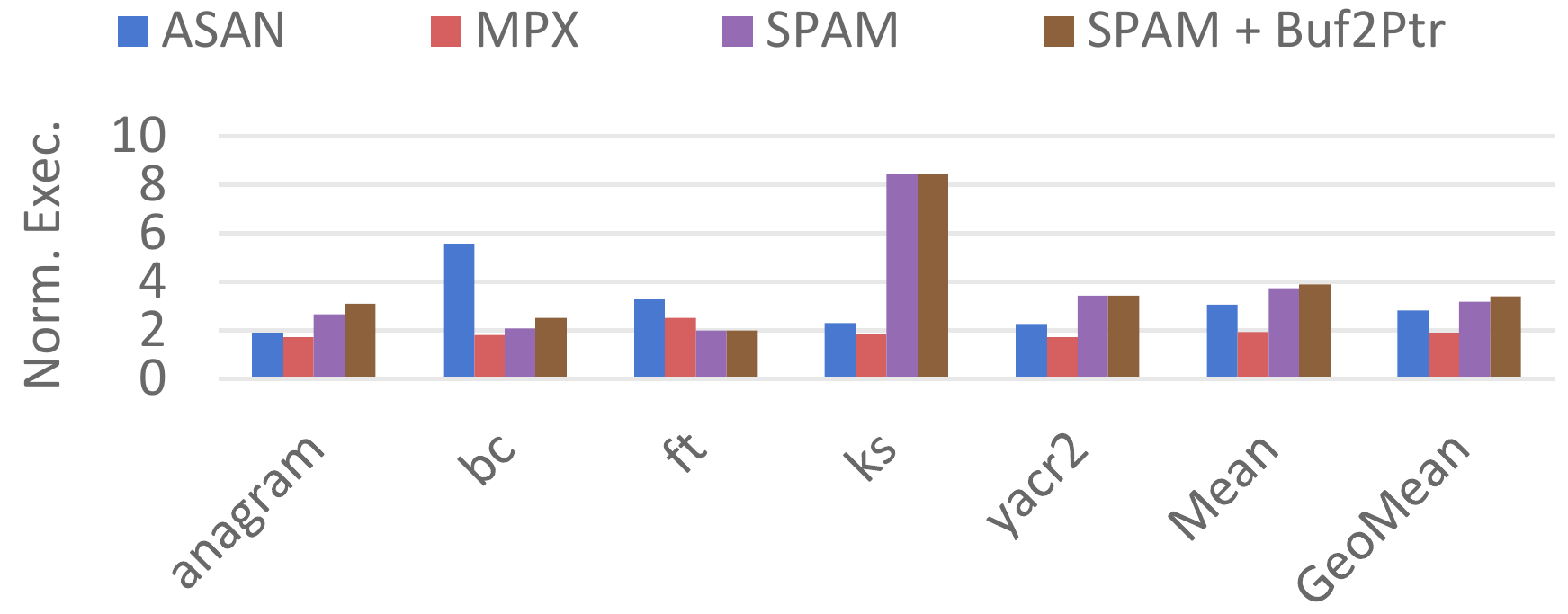}
  \caption{PtrDist Performance.}\label{fig:ptrdist-perf}
\end{figure}

\fakesec{PtrDist.} We evaluate~\pname{} performance compared to ASAN 
and Intel MPX using the Pointer-Intensive Benchmark
Suite~\cite{Todd1995:PtrDist}, or PtrDist, which is a collection of applications
specifically designed to test pointer-intensive operations. ~\cref{fig:ptrdist-perf} 
shows the evaluation results using the input set,
as specified by the \texttt{llvm-test-suite}.

\subsection{\pname{} Overheads Breakdown}

\begin{figure}[!h]
  \centering
  \includegraphics[width=0.45\textwidth]{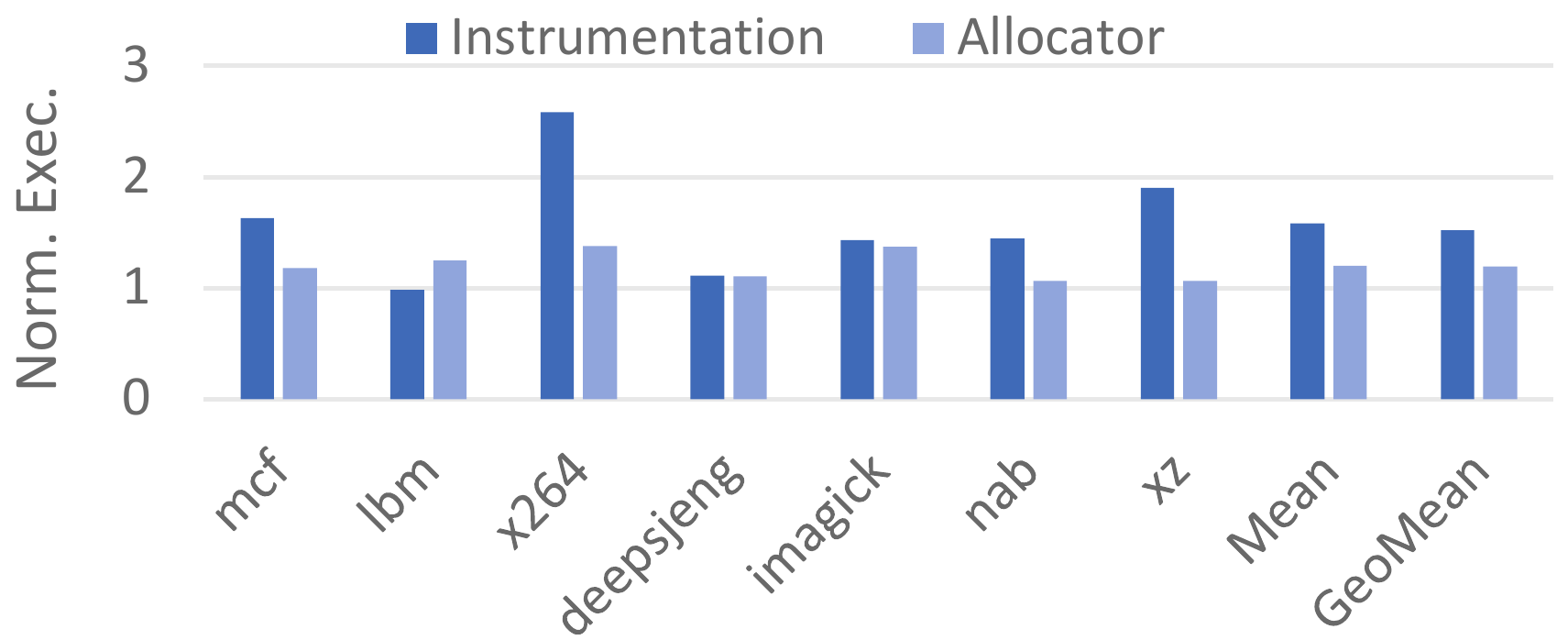}
  \caption{\pname{} overhead for SPEC2017.}\label{fig:spec-instrumentation}
\end{figure}

\cref{fig:spec-instrumentation} shows the performance overheads for \pname{}'s 
instrumentation and memory allocator without permuting program data for SPEC2017. 
The overheads are relative to the
uninstrumented baseline using the default system \texttt{malloc}.
Instrumentation (which includes the allocator overhead) on average accounts for $58\%$. 
There is further room for improvement by using more
accurate alias analysis, and writing tighter code for optimized case specific
functions allowing for more inlining opportunities. As we can see,
padding allocations to be multiples of 128 bytes adds an average allocator overhead of
$20\%$ for SPEC. While this allows permutations to be
computed more efficiently, it is worth noting that there may be room for future
performance gains by aligning to a smaller size.

\section{Library Wrappers}\label[appendix]{sec:opt-appendix}

To avoid paying the performance penalty of
unpermuting and permuting memory when making external library calls,
we implement \pname{} optimized wrappers for memory
intensive functions. This allows us to directly operate on permuted memory. The
most important of these functions are primarily found in the C standard
library's \incode{string.h} section (\eg{} \incode{memcpy}, \incode{memmove},
\incode{memset}, \incode{strcpy}, etc). \cref{tab:wrappers} lists the \texttt{libc} wrappers we
implemented to support the programs used in \cref{sec:evaluation}.

\begin{table}[!h]
\caption{\pname{} Wrapper List}\label{tab:wrappers}
\begin{tabularx}{0.45\textwidth}{Xl}
\tabitem \texttt{memcpy} & \tabitem \texttt{memmove}\\
\tabitem \texttt{memset} & \tabitem \texttt{memcmp}\\
\tabitem \texttt{memchr} & \tabitem \texttt{strtok}\\
\tabitem \texttt{strcmp} & \tabitem \texttt{strncmp}\\
\tabitem \texttt{strcasecmp} & \tabitem \texttt{strncasecmp}\\
\tabitem \texttt{strlen} & \tabitem \texttt{strnlen}\\
\tabitem \texttt{strcat} & \tabitem \texttt{strncat}\\
\tabitem \texttt{strspn} & \tabitem \texttt{strcspn}\\
\tabitem \texttt{strcpy} & \tabitem \texttt{strncpy}\\
\tabitem \texttt{strchr} & \tabitem \texttt{strstr}\\
\tabitem \texttt{strpbrk} & \tabitem \texttt{setvbuf}\\
\tabitem \texttt{setbuf} & \tabitem \texttt{setlinebuf}\\
\tabitem \texttt{setbuffer} & \tabitem \texttt{putenv}\\
\tabitem \texttt{getline} & \tabitem \texttt{getdelim}\\
\tabitem \texttt{realpath} & \tabitem \texttt{getcwd}\\
\tabitem \texttt{bcopy} & \tabitem \texttt{read}\\
\tabitem \texttt{write} & \tabitem \texttt{fopen}\\
\tabitem \texttt{fdopen} & \tabitem \texttt{freopen}\\
\tabitem \texttt{fread} & \tabitem \texttt{fwrite}\\
\tabitem \texttt{fclose} & \tabitem \texttt{opendir}\\
\tabitem \texttt{fdopendir} & \tabitem \texttt{closedir}\\
\tabitem \texttt{tmpfile} & \tabitem \texttt{writev} \\
\tabitem \texttt{epoll\_ctl} & \tabitem \texttt{epoll\_wait} \\
\end{tabularx}
\end{table}

\section{System-level Applicability}\label{sec:systemapp}

While our current implementation explores \pname{}'s applicability within the
application domain, we expand on how \pname{} integrates
in the context of an entire system.

\fakesec{OS Interface.} Only user-land applications are currently supported 
by our prototype. As a result, data needs to be unpermuted before making
any system-call (e.g., \texttt{write}). Similarly, we permute the data directly
after system calls that write to program memory (e.g., \texttt{read}). The above
data serialization adds additional runtime overheads and allows an attacker with
access to an OS vulnerability to corrupt the data while it is unpermuted. One
possible solution is to instrument OS APIs with~\pname{} so that no
serialization is needed. We leave this for future work.

\fakesec{Non~$64$-bit Architectures.} Non~$64$-bit systems (e.g., $32$-bit processors) are widely used in
Internet-of-Things and Cyber Physical Systems, not to mention a large body of
legacy systems. Thus, it is important to consider the applicability of \pname{}
to this class of devices. While our current implementation is not suited for
these systems due to our choice of LowFat allocator and use-after-free protection,
this does not preclude \pname{} from supporting~$32$-bit architectures.
To do so, we propose leveraging the flexibility to use a traditional allocator
(e.g., \texttt{dlmalloc}) while ensuring that no two C structs are of the same
size. The latter condition is to guarantee use-after-free protection in the
absence of the alias bits. As~\pname{} permutes structs based on base address
and size, having unique struct sizes guarantees unique permutations even if the
base address is the same.
\section{\pname{} Runtime APIs}\label[appendix]{sec:add-runtime-api}

We list the C API for the functionality described in Section~\ref{sec:imp}.

{\footnotesize
\begin{align}
  \mintinline{cpp}{void *GetBasePtr(void *Ptr)}
\end{align}
}%

\fakesec{Get Allocation Base Address.} \pname{} needs a way to retrieve the base address
of an allocation from an arbitrary pointer in order to then compute the
correct permutation.
How this metadata is retrieved varies depending on the
underlying allocator used to implement \pname{}. Our current prototype relies on the
LowFat allocator~\cite{Duck2016:LowFatSW} which is able to retrieve this information implicitly in
constant time by partitioning the memory space into aligned fixed sized
regions. It is important to note that broader \pname{} technique is not bound
to a specific allocator.

{\footnotesize
  \begin{align}
    \mintinline{cpp}{uint64_t GetSize(void *BaseAddr)}
  \end{align}
}%

\fakesec{Get Allocation Size.} In addition to the base address above, the
runtime also needs a mechanism to determine the bounds, or size, of an
allocation. Given the base address the underlying allocator must be able to
return this information. In our current implementation using LowFat, this information is
implicitly derived from the base address itself.

{\footnotesize
\begin{align}
  \mintinline{cpp}{uint64_t GenPerm(uint64_t K, void *BA, size_t S)}
\end{align}
}%

{\footnotesize
\begin{align}
\mintinline{cpp}{void *GetPermPtr(void *Ptr, uint64_t Perm)}
\end{align}
 }%

{\footnotesize
  \begin{equation}
    \begin{aligned}
      \mintinline{cpp}{void *Unpermute(void *Ptr)}\\
      \mintinline{cpp}{void *Permute(void *Ptr)}
    \end{aligned}
  \end{equation}
}%

{\footnotesize
  \begin{align}
      \mintinline{cpp}{void RegisterGlobal(void *Ptr)}
  \end{align}
}%

{\footnotesize
  \begin{align}
      \mintinline{cpp}{void *RegisterStack(void *Ptr)}
  \end{align}
}%

\end{document}